\documentclass[12pt]{article}   
   
\usepackage{array}   
\usepackage{epsfig}   
\usepackage{amssymb}   
\usepackage{graphics,graphpap}   
\usepackage{amssymb}   
\usepackage{amsmath}   
\usepackage[usenames]{color}  
\usepackage{slashed}   
\usepackage{graphicx}

\renewcommand{\thefootnote}{\fnsymbol{footnote}}   
   
\def\s#1{\setbox0=\hbox{$#1$}%   
  \rlap{\ifdim\wd0>.7em\kern.22\wd0\else\kern.1\wd0\fi /}#1}   

   
\begin{document}   
   
%%%%%%%%%% Title page   
\begin{titlepage}   
\begin{flushright}   
\begin{tabular}{l}   
IPPP-14-46\\ DCPT-14-92 
\end{tabular}   
\end{flushright}   
\vskip1.5cm   
\begin{center}   
{\Large \bf \boldmath   
Lifetimes and HQE}

\vskip1.3cm    
{\sc   
Alexander Lenz    \footnote{Alexander.Lenz@durham.ac.uk}$^{,1}$,

}   
\vskip0.5cm   
$^1$ Institute for Particle Physics Phenomenology (IPPP), 
      Department of Physics, Durham University, DH1 3LE, United Kingdom

\vskip2cm

%{\em Version of \today}   
   
\vskip3cm   
   
{\large\bf Abstract\\[10pt]} \parbox[t]{\textwidth}{   
 Kolya Uraltsev was one of the inventors of the Heavy Quark Expansion (HQE),
that  describes inclusive weak decays of hadrons containing heavy quarks and in 
particular lifetimes. 
Besides giving a pedagogic introduction into the subject, we review the development 
and the current 
status of the HQE, which just recently passed several non-trivial experimental 
tests with an unprecedented precision. In view of many new experimental results
for lifetimes of heavy hadrons, we also update several theory predictions:
$\tau (B^+)      / \tau (B_d)   = 1.04^{+0.05}_{-0.01} \pm 0.02 \pm 0.01$,
$\tau (B_s)      / \tau (B_d)   = 1.001 \pm 0.002$,
$\tau (\Lambda_b)/ \tau (B_d)   = 0.935 \pm 0.054$
and
$\bar {\tau} (\Xi_b^0)    / \bar{\tau} (\Xi_b^+) = 0.95 \pm 0.06$.
The theoretical precision is currently strongly limited by the unknown size of the 
non-perturbative matrix elements of four-quark operators, which could be determined
with lattice simulations.
}
   
\vfill   
   
%{\em submitted to JHEP}   
\end{center}   
\end{titlepage}   
   
\setcounter{footnote}{0}   
\renewcommand{\thefootnote}{\arabic{footnote}}   
\renewcommand{\theequation}{\arabic{section}.\arabic{equation}}   
   
\newpage

\section{Introduction}
\label{lifetime_sec1}
Lifetimes are among the most fundamental properties of elementary particles.
In this work we consider lifetimes of hadrons containing heavy quarks,
which decay via the weak interaction.
Their  masses and lifetimes read (according to PDG \cite{PDG}
and HFAG \cite{HFAG})
%m and tau_D: PDG
%      tau_B: HFAG

\vspace{0.4cm}

\centerline{\bf $B$-mesons}

\vspace{-0.1cm}

\begin{equation}
\begin{array}{|c||c|c|c|c|}
\hline
             &                  &                 &                  &                    
\\
             & B_d = (\bar{b}d) & B^+ = (\bar{b}u)& B_s = (\bar{b}s) & B^+_c = (\bar{b}c) 
\\
\hline
\mbox{Mass (GeV)}    &5.27955(26)& 5.27925(26)    & 5.3667(4)        & 6.2745(18)   
\\
\mbox{Lifetime (ps)} & 1.519(5)  & 1.638(4)       & 1.512(7)         & 0.500(13)  
\\
\tau (X) / \tau(B_d) & 1         & 1.076 \pm 0.004 & 0.995 \pm 0.006 & 0.329 \pm 0.009 
\\
\hline
\end{array}
\label{tauBexp}
\end{equation}

\vspace{0.4cm}

\centerline{\bf $b$-baryons}

\vspace{-0.1cm}

\begin{equation}
\begin{array}{|c||c|c|c|c|}
\hline
             &                  &                 &                  &                   
\\
             & \Lambda_b = (udb)& \Xi_b^0 = (usb) & \Xi_b^-= (dsb)   & \Omega_b^- = (ssb)
\\
\hline
\mbox{Mass (GeV)}    & 5.6194(6)&5.7918(5)       &5.79772(55)        & 6.071(40)
\\
\mbox{Lifetime (ps)} & 1.451(13)&1.477(32)       &1.599(46)          & 1.54 \left( {+26}\atop{-22} \right)
\\
\tau (X) / \tau(B_d) & 0.955 \pm 0.009& 0.972\pm0.021&1.053\pm0.030  & 1.01\left( {+17}\atop{-14} \right)
\\
\hline
\end{array}
\label{taubbaryexp}
\end{equation}
The masses and the lifetimes of the $\Xi_b^0$, $\Xi_b^-$  and the $\Omega_b^-$
have been measured by the LHCb Collaboration 
\cite{Aaij:2014esa, Aaij:2014lxa,Aaij:2014sia} 
just after the first version of this article appeared on the arXiv. 
We have given above these new values 
instead of the HFAG and PDG averages.
Alternative lifetime averages were, e.g., obtained in \cite{Stone:2014pra}.

\vspace{0.4cm}

\centerline{\bf $D$-mesons}

\begin{equation}
\begin{array}{|c||c|c|c|}
\hline
                    &                  &                 &                  
\\
                    & D^0 = (\bar{u}c) & D^+ = (\bar{d}c)& D_s^+ = (\bar{s}c) 
\\
\hline
\mbox{Mass (GeV)}   & 1.86491(17)      &  1.8695(4)    & 1.9690(14) 
\\
\mbox{Lifetime (ps)}& 0.4101(15)       & 1.040  ( 7)     & 0.500  (7)  
\\
\tau (X) / \tau(D^0)& 1                & 2.536 \pm 0.017 & 1.219  \pm 0.017  
\\
\hline
\end{array}
\label{tauDexp}
\end{equation}

\vspace{0.5cm}
\centerline{\bf $c$-baryons}

\begin{equation}
\begin{array}{|c||c|c|c|c|}
\hline
                    &               &                &                &
\\
                    &\Lambda_c=(udc)&\Xi_c^+ = (usc) &\Xi_c^0 = (dsc) &\Omega_c = (ssc) 
\\
\hline
\mbox{Mass (GeV)}   & 2.28646(14)   & 2.4676 \left( {+4}\atop{-10} \right) &
2.47109 \left( {+35}\atop{-100} \right) & 2.6952 \left( {+18}\atop{-16} \right)      
\\
\mbox{Lifetime (ps)}& 0.200(6)      & 0.442(26)&0.112 \left( {+13}\atop{-10} \right) &0.069(12)       
\\ 
\tau (X) / \tau(D^0)&0.488\pm 0.015 &   1.08(6) & 0.27(3)   & 0.17 \pm 0.03    
\\
\hline
\end{array}
\label{taucbaryexp}
\end{equation}

\vspace{0.3cm}

\noindent
One of the first observations to make is the fact that all lifetimes are of the same order
of magnitude,
they are all in the pico-second range and they differ at most by a factor of 25. 
Looking exclusively at hadrons containing one $b$-quark (and no $c$-quark), 
one even finds that all 
lifetimes are equal within about $10 \%$. This clearly calls for a theoretical explanation.
\\
In this review we will discuss the theoretical framework describing decay rates of
inclusive decays of hadrons containing a heavy quark, the so-called
{\bf H}eavy {\bf Q}uark {\bf E}xpansion. A special case of such observables are the lifetimes 
of hadrons, which are given by the inverse of the
total decay rates. Kolya Uraltsev was one of the main
pioneers in the development of the HQE, which has its roots back in the 1970s.
When I began my career, Kolya's theory was already a kind of textbook knowledge
and my PhD and my first scientific papers were devoted to the calculation of higher order 
QCD corrections
within the framework of the HQE. Thus it was very inspiring to meet Kolya personally at one of
my first international conferences, which was held in 2000 in Durham, where we discussed 
the so-called ``missing charm puzzle'' \cite{Lenz:2000kv} and the decay rate difference 
$\Delta \Gamma_s$
of $B_s$-mesons \cite{Beneke:2000cu}.
I benefited a lot from many follow-up encounters with Kolya, e.g., at CERN, in Portoroz and 
in Siegen. At the end of 2012 I was working with a student from Munich \cite{Lenz:2013aua} 
on $D$-meson lifetimes
and in that respect Kolya was sending me several long emails regarding 
the history of lifetime predictions, which clearly influenced this review.
\\
Many of the most convincing precision tests of the HQE have just been performed 
recently. In the beginning of 2012 $\Delta \Gamma_s$ has been measured 
for the first time, i.e., with a 
statistical significance of five standard deviations, in accordance with the 
HQE prediction. The 
long standing puzzle concerning the lifetime of the $\Lambda_b$-baryon - for many years
a very strong challenge for the HQE - seems to have been settled experimentally, with
the latest results just appearing in 2014.
It is a real tragedy that Kolya did not have more time to celebrate the 
successes of the theory, to which he contributed so much.
\\
We start in Section \ref{lifetime_sec2} with a very basic and pedagogic introduction
into lifetimes of weakly decaying particles. 
Readers with some familiarity to weak decays 
can skip Section 2.1, which covers some simple but instructive estimates
for the decays of the muon, the tauon, the charm-quark and the bottom-quark.
In Section 2.2 we discuss the structure of the HQE in detail and in Section 2.3
we give a brief review of the discussed observables.
In Section \ref{lifetime_sec3} we investigate the history of the HQE and we highlight
Kolya's contribution, while we discuss the 
status quo in experiment and theory in Section \ref{lifetime_sec4}. Here we also give some
numerical updates of theory predictions for lifetime ratios.
In Section \ref{lifetime_sec5} we give an outlook on what has to be done in order to improve
further the theoretical accuracy and we conclude.

\section{Basic considerations about lifetimes}
\label{lifetime_sec2}
We start the discussion of the theoretical framework for describing heavy hadron lifetimes,
with some simple estimates that are based on the prime example for a weak decay, the
decay of a muon. Later on we show how this naive picture has to be generalised for the case
of hadrons containing a heavy quark.

\subsection{Naive estimates}

\subsubsection{The muon decay}
The muon decay $\mu^-  \to \nu_\mu \; e^- \; \bar{\nu}_e$  represents the most 
simple weak decay, because there are no QCD effects involved.\footnote{This statements
 holds to a high accuracy. QCD effects arise for 
the first time at the two loop order.}
This process is given by the following Feynman diagram.
\begin{center}
\includegraphics[width = 0.5 \textwidth, angle = 0]{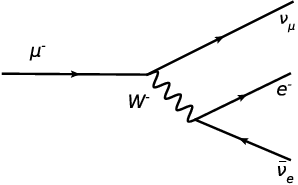}
\end{center}
Hence the total decay rate of the muon reads  (see, e.g., \cite{Michel:1949qe} 
for an early reference)
\begin{equation}
\Gamma_{\mu \to \nu_\mu + e + \bar{\nu}_e} =
\frac{G_F^2 m_\mu^5}{192 \pi^3} f \left(\frac{m_e}{m_\mu} \right) 
=
\frac{G_F^2 m_\mu^5}{192 \pi^3} c_{3,\mu}
\, .
\label{muon}
\end{equation}
$f$ denotes the phase space factor for one massive particle in the final state. It is
given by
\begin{eqnarray} 
f(x) & = & 1 - 8 x^2 + 8 x^6 - x^8 - 24 x^4 \ln (x)  \; .
\label{phase}
\end{eqnarray}
\noindent
The coefficient $c_{3, \mu}$ is introduced here to be consistent with our later notation.
The result in Eq.(\ref{muon}) is already very instructive, since we get now for
the measurable lifetime of the muon
\begin{equation}
      \tau = \frac{1}{\Gamma} = 
\frac{192 \pi^3}{G_F^2 m_\mu^5 f \left(\frac{m_e}{m_\mu}\right)} \; .
 \end{equation}
Thus the lifetime of a weakly decaying particle is proportional to the 
inverse of the fifth power of the mass of the decaying particle.
Using the measured values \cite{PDG} for 
$G_F=  1.1663787(6) \cdot 10^{-5} \; \mbox{GeV}^{-2}$ 
,
$m_e = 0.510998928 (11) $ MeV
and 
$m_\mu = 0.1056583715(35) $ GeV we predict\footnote{This is of course not 
really correct, because the measured
muon lifetime was used to determine the Fermi constant, but for pedagogical reasons
we assume that the Fermi constant is known from somewhere else.}
the lifetime of the muon to be
      \begin{equation}
       \tau_\mu^{Theo.} = 2.18776 \cdot 10^{-6} \, \mbox{s} \; ,
      \end{equation}
      which is in excellent agreement with the measured value \cite{PDG} of
      \begin{equation}
       \tau_\mu^{Exp.} = 2.1969811(22) \cdot 10^{-6} \, \mbox{s} \; .
      \label{muonexp}
      \end{equation}
      The remaining tiny difference (the prediction is about $0.4\% $ smaller than
      the experimental value) is due to higher order electro-weak corrections.
      These corrections are crucial for a high precision determination of the Fermi constant.
      The dominant contribution is given by the 1-loop QED correction, calculated already
      in the 1950s \cite{Behrends:1955mb,Kinoshita:1958ru}:
\begin{equation}
      c_{3, \mu} = f \left(\frac{m_e}{m_\mu}\right)
                \left[ 1 + \frac{\alpha}{4 \pi} 2 \left( \frac{25}{4} - \pi^2 \right)\right]\; .
\label{murad}
 \end{equation}
      Taking this effect into account ($\alpha =  1/137.035999074(44)$ \cite{PDG}) we predict
      \begin{equation}
       \tau_\mu^{Theo.} =  2.19699 \cdot 10^{-6} \, \mbox{s} \; ,
      \end{equation}
      which is almost identical to the measured value given in Eq.(\ref{muonexp}).
      The complete 2-loop QED corrections have been determined in \cite{vanRitbergen:1998yd}, 
      a review of loop-corrections to the muon decay is given in
      \cite{Sirlin:2012mh} and two very recent higher order calculations can be found in, e.g.,
      \cite{Ferroglia:2013dga,Fael:2013pja}.
      \\
      The phase space factor is almost negligible for 
      the muon decay - it reads $f(m_e/m_\mu) = 0.999813 = 1- 0.000187051$ - 
      but it will turn out to be quite sizable for a decay of a b-quark into a charm quark.
\subsubsection{The tau decay}
      Moving to the tau lepton, 
      we have now two leptonic decay channels as well as decays into quarks:
      \begin{eqnarray}
      \tau & \to & \nu_\tau + 
        \left\{ \begin{array}{l}
                e^- + \bar{\nu}_e 
                \\
                \mu^- + \bar{\nu}_\mu 
                \\
                d + \bar{u} 
                \\
                s + \bar{u} 
                \end{array}
        \right. \;.
      \nonumber 
      \end{eqnarray}
\begin{center}
\includegraphics[width = 0.49 \textwidth, angle = 0]{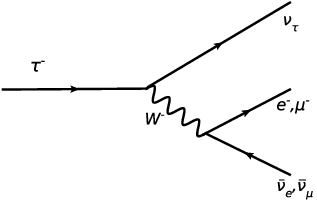}
\hfill
\includegraphics[width = 0.49 \textwidth, angle = 0]{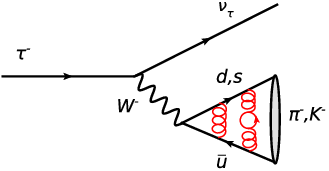}
\end{center}

\noindent
      Heavier quarks, like charm- or bottom-quarks cannot be created, because the lightest meson
      containing such quarks ($D^0 = c \bar{u}; M_{D^0} \approx 1.86 \, \mbox{GeV})$ is heavier 
      than the tau lepton $(m_\tau = 1.77682(16) \; \mbox{GeV})$.
      Thus the total decay rate of the tau lepton reads     
      \begin{eqnarray}
      \Gamma_\tau & = &
       \frac{G_F^2 m_\tau^5}{192 \pi^3} \left[
                               f \left(\frac{m_e  }{m_\tau} \right) 
       +                       f \left(\frac{m_\mu}{m_\tau} \right) 
       + N_c \left|V_{ud}\right|^2 g \left(\frac{m_u}{m_\tau},\frac{m_d}{m_\tau} \right) 
       + N_c \left|V_{us}\right|^2 g \left(\frac{m_u}{m_\tau},\frac{m_s}{m_\tau} \right) 
\right]
\nonumber
\\
& =: &
 \frac{G_F^2 m_\tau^5}{192 \pi^3}  c_{3, \tau}
\, .
\end{eqnarray}
The factor $N_c=3$ is a colour factor
and $g$ denotes a new phase space function, when there are two massive particles 
in the final state. 
If we neglect the phase space factors 
($f(m_e  / m_\tau) = 1-7\cdot 10^{-7}; 
f(m_\mu/ m_\tau) = 1-0.027;...$) and  if we use
$V_{ud}^2 + V_{us}^2 \approx 1$, then we get $c_{3,\tau} = 5$ and thus the simple approximate relation
\begin{equation}
\frac{\tau_\tau}{\tau_\mu} = \left( \frac{m_\mu}{m_\tau}\right)^5 \frac{1}{5} \; .
\end{equation}
Using the experimental values for $\tau_\mu$, $m_\mu$ and $m_\tau$ we predict
\begin{equation}
\tau_\tau^{Theo.} = 3.26707 \cdot 10^{-13} \; \mbox{s} \; ,
\end{equation}
which is quite close to the experimental value of
\begin{equation}
\tau_\tau^{Exp.} = 2.906(1) \cdot 10^{-13} \; \mbox{s} \; .
\end{equation} 
Now the theory prediction is about $12 \%$ larger than the
measured value. This is mostly due to sizable QCD corrections, when
there are quarks in the final state - which was not possible in the
muon decay. 
These QCD corrections are currently 
calculated up to five loop accuracy 
\cite{Baikov:2008jh}, a review of higher order corrections
can be found in \cite{Pich:2013lsa}.
\\
Because of the pronounced and clean dependence on the strong coupling, tau decays can 
also be used for precision determinations of $\alpha_s$, see, e.g., the review
\cite{Altarelli:2013bpa}.
This example shows already, that a proper treatment of QCD effects is mandatory for 
precision investigations of lifetimes. In the case of meson decays this will even be more important.

\subsubsection{Meson decays }
Studying  weak decays of mesons - instead of leptons - one immediately comes to the problem
of treating the QCD effects related to the formation of a bound state in the initial state as
well as in the final states and there  is of course also the possibility of having QCD effects 
between the initial and the final states.
Meson decays can be classified according to their final states:
\begin{itemize}
\item \underline{\bf Leptonic decays} have only leptons in the final state, 
      e.g., $B^- \to \tau^- \;  \bar{\nu}_\tau$.
      \begin{center}
        \centerline{\includegraphics[width=0.5\textwidth,angle = 0]{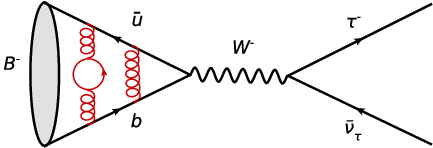}}
      \end{center}
      \vspace{-0.5cm}
      Such decays have the simplest hadronic structure. Gluons bind the quarks
      of the initial state into a hadron. All non-perturbative effects  
      are described by so-called decay constants.
\item \underline{\bf Semi-leptonic decays} have both leptons and hadrons in the final state, 
      e.g., $B^- \to D^0 \; e^-  \; \bar{\nu}_e$.
      \begin{center}
        \centerline{\includegraphics[width=0.5\textwidth,angle = 0]{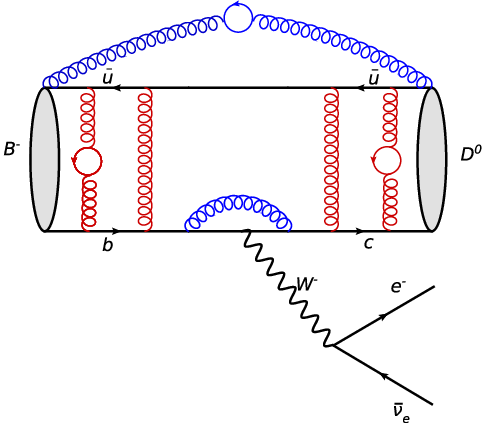}}
      \end{center}
      \vspace{-0.5cm}
      Now the hadronic structure is more complicated. We have the binding of hadrons in the
      initial state and in the final states. Moreover there is the possibility of having
      strong interactions between the initial state and the final states. The non-perturbative
      physics is in this case described by so-called form factors.
\item \underline{\bf Non-leptonic decays} have only hadrons in the final state, 
      e.g., $B^- \to  D^0 \; \pi^-$.
      \begin{center}
        \centerline{\includegraphics[width=0.5\textwidth,angle = 0]{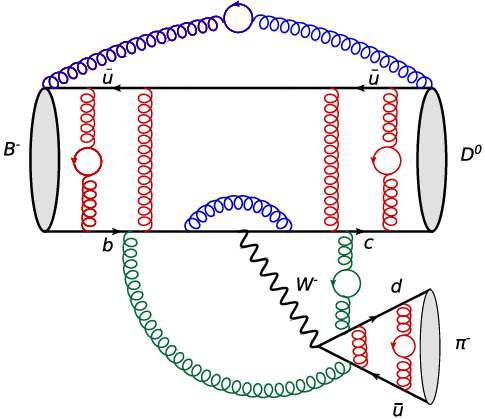}}
      \end{center}
      \vspace{-0.7cm}
      These are the most complicated decays and they can only be treated by making additional
      assumptions that allow then for a factorisation. 
\end{itemize}
Related to this considerations we introduce the notation for two classes of decays 
- {\it inclusive} and {\it exclusive} decays.
In exclusive modes every final state hadron is identified. This is in principle what 
experiments can do well, while theory has the
problem to describe the peculiar QCD binding effects in the hadronic states.
An example from above would be the decay $ B^- \to  D^0 \; \pi^-$, where one explicitly
detects the $D^0$ and $\pi^-$ in the final state. The corresponding inclusive
decay is $ b \to c \bar{u} d$.
In inclusive modes we only care about the quark content of the final state,
this is clearly theoretically easier, while experiments have the problem
of summing up all decays that belong to a certain inclusive decay mode.
Another example that will appear later on, is the  inclusive 
 $b \to c \; \bar{c}  \; s$ transition; corresponding exclusive decays are
in this case
 \begin{eqnarray}
 B_d^0 & \to &  D^{*-} \; D_s^{*+} \, ,
% \nonumber
% \\
%       & \to &
  D^-    \; D_s^{*+}  \, ,
% \nonumber
% \\
%       & \to &  
D^{*-} \; D_s^{+}\, ,
% \nonumber
% \\
%       & \to &  
D^-    \; D_s^{+} \, ,
% \nonumber
% \\
%      & \to & 
J/\Psi   \; K_S     \; ,
% \nonumber
% \\
%      & \to & 
\dots \; .
 \nonumber
 \end{eqnarray}
To get a feeling for the arising branching fractions we list the 
theory value \cite{Krinner:2013cja} for
 $b \to c \; \bar{c}  \; s$, with some measured \cite{PDG} 
exclusive branching ratios.
\begin{eqnarray}
{\rm Br} (b \to c \bar{c}  s) & = & (23 \pm 2) \% \; ,
\\
{\rm Br} (D^{*-} \; D_s^{*+}) & = & (1.77 \pm 0.14) \% \; ,
\\
{\rm Br} (D^{*-} \; D_s^{+})  & = & (8.0 \pm 1.1) \cdot 10^{-3} \; ,
\\
{\rm Br} (D^{-} \; D_s^{*+})  & = & (7.4 \pm 1.6) \cdot 10^{-3} \; ,
\\
{\rm Br} (D^{-} \; D_s^{+})   & = & (7.2 \pm 0.8) \cdot 10^{-3} \; ,
\\
{\rm Br} (J/\Psi \; K_S)      & = & (8.73 \pm 0.32) \cdot 10^{-4} \; .
\end{eqnarray}
Here one can already guess that quite some number of exclusive decay channels has to be 
summed up in order to obtain the inclusive branching ratio.
\subsubsection{Charm-quark decay}
Before trying to investigate the complicated meson decays, let us look at 
the decay of free $c$- and $b$-quarks.
Later on  we will show that the free quark decay is the
leading term in a systematic expansion in the inverse of the 
heavy (decaying) quark mass - the HQE.
\\
A charm quark can decay weakly into a strange- or a down-quark and 
a $W^+$-boson, which then further decays either into leptons (semi-leptonic decay)
or into quarks (non-leptonic decay).
\begin{center}
\includegraphics[width = 0.49 \textwidth, angle = 0]{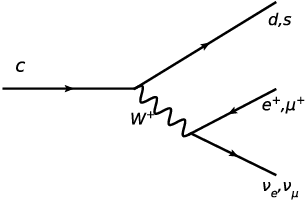}
\hfill
\includegraphics[width = 0.49 \textwidth, angle = 0]{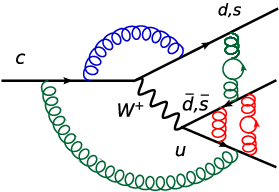}
\end{center}
Calculating the total inclusive decay rate of a charm-quark we get
\begin{eqnarray}
\Gamma_c =
\frac{G_F^2 m_c^5}{192 \pi^3} |V_{cs}|^2 c_{3,c} \; ,
\end{eqnarray}
with
\begin{eqnarray}
c_{3,c} = && 
            g\left(\frac{m_s}{m_c},\frac{m_{e}  }{m_c} \right)
                +  g\left(\frac{m_s}{m_c},\frac{m_{\mu}}{m_c} \right)
                + N_c |V_{ud}|^2 h\left(\frac{m_s}{m_c}, \frac{m_u}{m_c}, \frac{m_d}{m_c} \right)
                + N_c |V_{us}|^2 h\left(\frac{m_s}{m_c}, \frac{m_u}{m_c}, \frac{m_s}{m_c} \right)
\nonumber
\\
&& 
+
\left|\frac{V_{cd}}{V_{cs}}\right|^2 
           \left\{g\left(\frac{m_d}{m_c},\frac{m_e  }{m_c}\right)
                + g\left(\frac{m_d}{m_c},\frac{m_\mu}{m_c}\right)
                + N_c |V_{ud}|^2 h\left(\frac{m_d}{m_c}, \frac{m_u}{m_c}, \frac{m_d}{m_c} \right)
\right. 
\nonumber
\\
&& 
\left. 
\hspace{6.3cm}    + N_c |V_{us}|^2 h\left(\frac{m_d}{m_c}, \frac{m_u}{m_c}, \frac{m_s}{m_c} \right)
                    \right\} \; .
\end{eqnarray}
$h$ denotes a new phase space function, when there are three massive particles 
in the final state. If we set all phase space factors to one 
($f(m_s/m_c) = f(0.0935/1.471) = 1 - 0.03, \dots $ with $m_s = 93.5(2.5) $ MeV \cite{PDG})
and use $|V_{ud}|^2+|V_{us}|^2 \approx 1 \approx |V_{cd}|^2+|V_{cs}|^2$, then we get
$ c_{3,c} = 5$, similar to the $\tau$ decay. In that case  
we predict  a charm lifetime of
\begin{eqnarray}
\tau_c & = & \left\{ \begin{array}{l}
                    0.84 \; \mbox{ps}
                    \\
                    1.70 \; \mbox{ps}
                    \end{array} \right.
             \hspace{0.5cm} \mbox{for} \; m_c =
                  \left\{ \begin{array}{ll}
                   1.471   &\mbox{GeV} \; \; (\mbox{Pole-scheme})
                    \\
                   1.277(26) & \mbox{GeV} \; \; (\overline{MS}-\mbox{scheme})
                    \end{array} \right. \; .
\label{tauc}
\end{eqnarray}
These predictions lie roughly  in the ball-bark of the experimental numbers for $D$-meson 
lifetimes, but at this stage some comments are appropriate:
\begin{itemize}
\item Predictions of the lifetimes of free quarks have a huge parametric dependence 
      on the definition of the quark mass ($\propto m_q^5$). This is the reason, why typically
      only lifetime ratios (the dominant $m_q^5$ dependence as well as CKM factors and some
      sub-leading non-perturbative corrections cancel) are determined theoretically. We show 
      in this introduction for pedagogical reasons the numerical results of the theory
      predictions of lifetimes and not only ratios. 
      In our case the value obtained with the $\overline{MS}-\mbox{scheme}$
      for the charm quark mass is about a factor of 2 larger than the one obtained with
      the pole-scheme. In LO-QCD the definition of the quark mass is completely arbitrary
      and we have these huge uncertainties. If we calculate everything consistently
      in NLO-QCD, the treatment of the quark masses has to be defined within the calculation,
      leading to a considerably weaker dependence of the final result on the quark mass 
      definition.
      \\
      Bigi, Shifman, Uraltsev and Vainshtein have shown in 1994 \cite{Bigi:1994em} that the pole
      mass scheme is always affected by infra-red renormalons, see also the paper of Beneke
      and Braun \cite{Beneke:1994sw} that appeared on the same day on the arXiv 
      and the review in this issue \cite{Shifman:2013uka}. Thus 
      short-distance definitions of the quark mass, like the $\overline{\rm MS}$-mass 
      \cite{Bardeen:1978yd} seem to be better suited than the pole mass. More recent 
      suggestions for quark mass concepts are the kinetic mass from Bigi, Shifman, 
      Uraltsev and Vainshtein \cite{Bigi:1994ga,Bigi:1996si} introduced in 1994, 
      the potential subtracted mass from 
      Beneke \cite{Beneke:1998rk} 
      and the $\Upsilon(1s)$-scheme from Hoang, Ligeti and Manohar \cite{Hoang:1998ng,Hoang:1998hm}, both
      introduced in 1998.
      In \cite{Krinner:2013cja} we compared the above quark mass schemes for 
      inclusive non-leptonic decay rates and found similar numerical results for the different
      short distance masses.
      Thus we rely in this review - for simplicity - on predictions based on the 
      $\overline{\rm MS}$-mass scheme 
      and we discard the pole mass, even if we give several times predictions
      based on this mass scheme for comparison.
      \\
      Concerning the concrete numerical values for the quark masses we also take the same 
      numbers as in \cite{Krinner:2013cja}.
      In that work relations between different quark mass schemes were strictly used at NLO-QCD
      accuracy (higher terms were discarded), therefore the numbers differ slightly from 
      the PDG \cite{PDG}-values, which 
      would result in
      \begin{eqnarray}
      \tau_c & = & \left\{ \begin{array}{l}
                    0.44 \; \mbox{ps}
                    \\
                    1.71 \; \mbox{ps}
                    \end{array} \right.
             \hspace{0.5cm} \mbox{for} \; m_c =
                  \left\{ \begin{array}{ll}
                   1.67(7)   &\mbox{GeV} \; \; (\mbox{Pole-scheme})
                    \\
                   1.275(25) & \mbox{GeV} \; \; (\overline{MS}-\mbox{scheme})
                    \end{array} \right. \; .
      \end{eqnarray}
      Since our final lifetime predictions are only known up to NLO accuracy and  we expand
      every expression consistently up to order $\alpha_s$, we will stay with the parameters
      used in  \cite{Krinner:2013cja}.
\item Taking only the decay of the $c$-quark into account, one obtains the same lifetimes
      for all charm-mesons, which is clearly a very bad approximation, taking the large spread
      of lifetimes of different $D$-mesons into account, see Eq.(\ref{tauDexp}).
      Below we will see that in the case of charmed mesons a very sizable contribution
      comes from non-spectator effects where also the valence quark of the $D$-meson 
      is involved in the decay.
\item Perturbative QCD corrections will turn out to be very important, because 
      $\alpha_s(m_c)$ is quite large.
\item In the above expressions we neglected, e.g., 
      annihilation decays like $D^+ \to l^+ \; \nu_l$, 
      which have very small branching ratios \cite{PDG} (the corresponding Feynman diagrams 
      have the same topology as 
      the decay $B^- \to \tau^- \bar{\nu}_\tau$, that was mentioned earlier). 
      In the case of $D_s^+$ meson the branching ratio
      into $\tau^+ \; \nu_\tau$ will, however,  be sizable \cite{PDG} and has to be taken 
      into account.
      \begin{equation}
      {\rm Br} (D_s^+ \to \tau^+ \; \nu_\tau) = (5.43 \pm 0.31) \% \; .
      \end{equation}
\end{itemize}
In the framework of the HQE the non-spectator effects will turn out to be suppressed 
by $1/m_c$ and since 
$m_c$ is not very large, the suppression is also not expected to be very pronounced. 
This will change in the case of $B$-mesons. Because of the larger value of the 
$b$-quark mass, one expects a better description of the meson decay in terms of  the simple
$b$-quark decay.
\subsubsection{Bottom-quark decay}
\begin{center}
\includegraphics[width = 0.49 \textwidth, angle = 0]{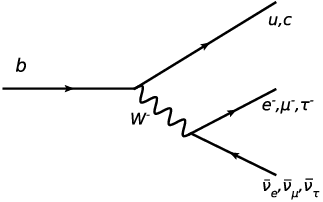}
\hfill
\includegraphics[width = 0.49 \textwidth, angle = 0]{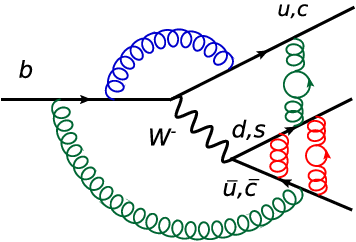}
\end{center}
Calculating the total inclusive decay rate of a $b$-quark we get
\begin{eqnarray}
\Gamma_b =
\frac{G_F^2 m_b^5}{192 \pi^3}|V_{cb}|^2 c_{3,b} \; ,
\end{eqnarray}
with
\begin{eqnarray}
c_{3, b} = && 
 \left\{g\left(\frac{m_c}{m_b}, \frac{m_e   }{m_b}\right)
                + g\left(\frac{m_c}{m_b}, \frac{m_\mu }{m_b}\right)
                + g\left(\frac{m_c}{m_b}, \frac{m_\tau}{m_b}\right)
\right.
\nonumber
\\
&& %\hspace{1.5cm}
\left.
\; \; \; \; \; \;  \; \; \; \; \;
             + N_c |V_{ud}|^2 h\left(\frac{m_c}{m_b},\frac{m_u}{m_b},\frac{m_d}{m_b} \right)
             + N_c |V_{us}|^2 h\left(\frac{m_c}{m_b},\frac{m_u}{m_b},\frac{m_s}{m_b} \right)
\right.
\nonumber
\\
&& %\hspace{1.5cm}
\left.
\; \; \; \; \; \;  \; \; \; \; \;
                + N_c |V_{cd}|^2 h\left(\frac{m_c}{m_b},\frac{m_c}{m_b},\frac{m_d}{m_b} \right)
                + N_c |V_{cs}|^2 h\left(\frac{m_c}{m_b},\frac{m_c}{m_b},\frac{m_s}{m_b} \right)
                    \right\}
\nonumber
\\
&& %\hspace{0.5cm}
+
\left| \frac{V_{ub}}{V_{cb}}\right|^2 \left\{g\left(\frac{m_u}{m_b}, \frac{m_e   }{m_b}\right)
                + g\left(\frac{m_u}{m_b}, \frac{m_\mu }{m_b}\right)
                + g\left(\frac{m_u}{m_b}, \frac{m_\tau}{m_b}\right)
\right.
\nonumber
\\
&& %\hspace{1.5cm}
\left. 
\; \; \; \; \; \;  \; \; \; \; \;
             + N_c |V_{ud}|^2 h\left(\frac{m_u}{m_b},\frac{m_u}{m_b},\frac{m_d}{m_b} \right)
             + N_c |V_{us}|^2 h\left(\frac{m_u}{m_b},\frac{m_u}{m_b},\frac{m_s}{m_b} \right)
\right.
\nonumber
\\
&& %\hspace{1.5cm}
\left.
\; \; \; \; \; \;  \; \; \; \; \;
                + N_c |V_{cd}|^2 h\left(\frac{m_u}{m_b},\frac{m_c}{m_b},\frac{m_d}{m_b} \right)
                + N_c |V_{cs}|^2 h\left(\frac{m_u}{m_b},\frac{m_c}{m_b},\frac{m_s}{m_b} \right)
                    \right\}
\; .
\nonumber
\\
\label{taubc}
\end{eqnarray}
In this formula penguin induced decays have been neglected, 
they will enhance the decay rate by several per cent, 
see \cite{Krinner:2013cja}. More important will, however, be the QCD corrections. 
To proceed further we can neglect
the masses of all final state particles, except for the charm-quark and for the tau lepton.
In addition we can neglect the contributions proportional to $|V_{ub}|^2$ since 
$|V_{ub}/V_{cb}|^2 \approx 0.01$. Using further  
$|V_{ud}|^2+|V_{us}|^2 \approx 1 \approx |V_{cd}|^2+|V_{cs}|^2$, 
we get the following simplified formula
\begin{eqnarray}
c_{3,b}
& = & \left[         (N_c + 2) f \left(\frac{m_c}{m_b}                    \right)
                  +   g \left(\frac{m_c}{m_b},\frac{m_\tau}{m_b} \right) 
                  + N_c g \left(\frac{m_c}{m_b},\frac{m_c}{m_b}     \right)
\right] \; .
\end{eqnarray}
If we have charm quarks in the final states, then the phase space functions show a 
huge dependence on the numerical value of the charm quark mass
(values taken from \cite{Krinner:2013cja})
\begin{equation}
f\left(\frac{m_c}{m_b}\right) =
\left\{ \begin{array}{l}
                     0.484
                    \\
                     0.518
                    \\
                     0.666
                    \end{array} \right.
\hspace{0.5cm} \mbox{for} 
                  \left\{ \begin{array}{llllll}
 m_c^{\rm Pole}      &= 1.471  &\mbox{GeV}, & m_b^{\rm Pole}       &= 4.650 &\mbox{GeV} 
                    \\
\bar{m}_c (\bar{m}_c)&= 1.277 &\mbox{GeV}, & \bar{m}_b (\bar{m}_b) &= 4.248 &\mbox{GeV} 
                    \\
\bar{m}_c (\bar{m}_b)&= 0.997 &\mbox{GeV}, & \bar{m}_b (\bar{m}_b) &= 4.248 &\mbox{GeV} 
                    \end{array} \right. \; .
\end{equation}
The big spread in the values for the space functions clearly shows again 
that the definition of the quark mass is a critical issue for a precise 
determination of lifetimes. The value for the pole quark mass is only shown to
visualise the strong mass dependence. As discussed above short-distance masses 
like the $\overline{\rm MS}$-mass are theoretically better suited.
Later on we will argue further for using $\bar{m}_c (\bar{m}_b)$ and $\bar{m}_b (\bar{m}_b)$
- so both masses at the scale $m_b$ -, 
which was suggested in \cite{Beneke:2002rj}, in order to sum up 
large logarithms of the form $\alpha_s^n (m_c/m_b)^2 \log^n (m_c/m_b)^2$ to all orders. 
Thus only the result using  $\bar{m}_c (\bar{m}_b)$ and $\bar{m}_b (\bar{m}_b)$ should be
considered as the theory prediction, while the additional numbers are just given for 
completeness.
\\
The phase space function for two identical particles in the final states reads 
\cite{Gourdin:1979qq,Blok:1992he,Bigi:1993fm,Bigi:1994wa}
(see \cite{Falk:1994gw} for the general case of three different masses)
\begin{equation}
g(x) = \sqrt{1-4x^2} \left(1-14 x^2 - 2 x^4 - 12 x^6\right) + 24 x^4 \left(1-x^4\right) 
        \log \frac{1+\sqrt{1-4 x^2}}{1-\sqrt{1-4 x^2}} \; ,
\label{phase2}
\end{equation}
with $x = m_c/m_b$.
Thus we get in total for all the phase space contributions
\begin{equation}
c_{3,b}
= \left\{ \begin{array}{l}
               9 \\ 2.97 \\  3.25 \\  4.66 \\ 
               \end{array}
       \right.
 \hspace{0.5cm} \mbox{for} \; \; 
 \left\{ \begin{array}{l}
               m_c= 0, 
               \\  
               m_c^{\rm Pole}, m_b^{\rm Pole}
               \\
               \bar{m}_c(\bar{m}_c),  \bar{m}_b(\bar{m}_b)
               \\
               \bar{m}_c(\bar{m}_b),  \bar{m}_b(\bar{m}_b)
               \end{array}
       \right.  \; .
\label{c3num}
\end{equation}
The phase space effects are  now quite dramatic.
For the total $b$-quark lifetime we predict 
(with $V_{cb} = 0.04151^{+0.00056}_{-0.00115}$ from \cite{Charles:2004jd}, 
for similar results see \cite{Ciuchini:2000de}.)
\begin{eqnarray}
\tau_b& = & \left. \begin{array}{l}
                    2.60 \; \mbox{ps}
                    \end{array} \right.
             \hspace{0.5cm} \mbox{for} 
                  \left. \begin{array}{ll}
\bar{m}_c(\bar{m}_b), &\bar{m}_b(\bar{m}_b)
                    \end{array} \right. \; .
\label{taubm0}
\end{eqnarray}
This number is about $70\%$ larger than the experimental number for the $B$-meson lifetimes.
There are in principle two sources for that discrepancy: first we neglected several 
CKM-suppressed decays,
which are  however not phase space suppressed as well as penguin decays. An inclusion of these
decays will enhance the total decay rate roughly by about $10\%$ and thus reduce the lifetime
prediction by about $10\%$. 
Second, there are large QCD effects,
that will be discussed in the next subsection;
including them will bring our theory prediction very close to the experimental number.
For completeness we show also the lifetime predictions, for different 
(theoretically less motivated) values of the quark masses.
\begin{eqnarray}
\tau_b& = & \left\{ \begin{array}{l}
                    0.90 \; \mbox{ps}
                    \\
                    1.42 \; \mbox{ps}
                    \\
                    2.59 \; \mbox{ps}
                    \\
                    3.72 \; \mbox{ps}
                    \end{array} \right.
             \hspace{0.5cm} \mbox{for} 
                  \left\{ \begin{array}{ll}
 m_c = 0 ,            & m_b^{\rm Pole}
\\
 m_c = 0 ,            & \bar{m}_b(\bar{m}_b) 
\\
 m_c^{\rm Pole} ,      & m_b^{\rm Pole}
\\
\bar{m}_c(\bar{m}_c), &\bar{m}_b(\bar{m}_b)
                    \end{array} \right. \; .
\label{taubm1}
\end{eqnarray}
By accident a neglect of the charm quark mass can lead to 
predictions that are very close to experiment. As argued above,
only the value in Eq.(\ref{taubm0}) should be considered as the theory
prediction for the $b$-quark lifetime and not the ones in Eq.(\ref{taubm1}).
Next we introduce the missing, but necessary concepts for making 
reliable predictions for the lifetimes of heavy hadrons.

\subsection{The structure of the HQE}

\subsubsection{The effective Hamiltonian}
Above we tried to make clear, that for any numerical reliable quantitative estimate 
of meson decays, QCD effects have to be taken 
properly into account.
To do so, weak decays of heavy quarks are not described within the full
standard model, but with the help of an effective Hamiltonian.
We start here simply with the explicit form of the effective Hamiltonian 
and refer the interested
reader to some excellent reviews by
Buchalla, Buras and Lautenbacher \cite{Buchalla:1995vs}, by Buras \cite{Buras:1998raa},
by Buchalla \cite{Buchalla:2002pd} and a recent one by Grozin \cite{Grozin:2013hra}.
The effective Hamiltonian reads
\begin{eqnarray}
\label{heff}
 {\cal H} _{eff}= \frac{G_{F}}{ \sqrt{2}} \left[ \sum_{q=u,c} V_{c}^q
 ( { C_{1}} { Q_{1}^q} + 
   { C_{2}} { Q_{2}^q} )-V_{p} \sum_{j=3} 
   { C_{j}} { Q_{j} } \right] \, .
\end{eqnarray}
Without QCD corrections only the operator $Q_2$ arises and the 
{\it Wilson coefficient} $C_2$ is equal to one, $C_2 = 1$. $Q_2$ 
has a current-current structure:
\begin{equation}
Q_2 = c_\alpha \gamma_\mu (1 - \gamma_5) \bar b_\alpha 
      \times 
      d_\beta  \gamma^\mu (1 - \gamma_5) \bar u_\beta
\, .
\end{equation}
$\alpha$ and $\beta$ denote colour indices.
The $V$s describe different combinations 
of CKM elements. With the inclusion of QCD one gets
additional operators. $Q_1$ has the same Dirac structure as $Q_2$, but it has a 
different colour structure, $Q_3,..,Q_6$ arise from QCD penguin diagrams etc..
Due to renormalisation all Wilson coefficients become scale 
dependent functions. In LO-QCD we get\footnote{We use as an input for the strong coupling
$a_s(M_Z) = 0.1184$.} $C_2(4.248 \, {\rm GeV}) = 1.1$ and
$C_1(4.248 \,{\rm GeV}) = -0.24$ and the penguin 
coefficients are below $5\%$, with the exception of $C_8$, the coefficient 
of the chromo-magnetic operator.
With this  operator product expansion (OPE) a separation of the scales was achieved. 
The high energy physics is described by the Wilson coefficients, they can be calculated in 
perturbation theory. The low energy physics is described by the matrix 
elements of the operators $Q_i $. Moreover large logarithms of the form 
$\alpha_s(m_b) \ln (m_b^2/M_W^2)$, which spoil the perturbative expansion in the full
standard model, are now summed up to all orders.
For semi-leptonic decays like $b \to c l^- \bar{\nu}_l$ the Wilson coefficient $C_2$ is simply
1, while the remaining ones vanish.

\subsubsection{The free quark decay with the effective Hamiltonian}
\label{Heff}
Now we can calculate the free quark decay starting from the effective Hamiltonian instead of the
full standard model. If we again neglect penguins, we get
in leading logarithmic approximation the same structure as in Eq.(\ref{taubc}),
\begin{eqnarray}
c_{3,b}^{\rm LO-QCD}
& = &  c_{3,b}^{c e \bar{\nu}_e} +  c_{3,b}^{c \mu \bar{\nu}_\mu} 
    +  c_{3,b}^{ c \bar{u} d}    +  c_{3,b}^{ c \bar{u} s}        
    +  c_{3,b}^{c \tau \bar{\nu}_\tau}
    +  c_{3,b}^{ c \bar{c} s}    +  c_{3,b}^{ c \bar{c} d} 
\dots 
\\
& = & \left[   \left( 2 + {\cal N}_a(\mu) \right)    f \left(\frac{m_c}{m_b}                    \right)
                  +   g \left(\frac{m_c}{m_b},\frac{m_\tau}{m_b} \right) 
                  +{\cal N}_a(\mu) g \left(\frac{m_c}{m_b},\frac{m_c}{m_b}     \right)
\right] ,
\label{c3LOQCD} 
\end{eqnarray}
with changing the colour factor $N_c=3$ - stemming from QCD -  into
\begin{eqnarray}
{\cal N}_a(\mu) & = & 
3 C_1^2(\mu) + 3 C_2^2(\mu) + 2 C_1(\mu) C_2(\mu) \approx 3.3 \; \; 
   ({\rm LO}, \; \mu = 4.248 \; {\rm GeV})
\; .
\end{eqnarray}
This effect enhances the total decay rate by about $10\%$ and thus brings down (if also the
sub-leading decays are included) the
prediction for the lifetime of the $b$-quark to about
\begin{eqnarray}
\tau_b& \approx & \left. \begin{array}{l}
                    2.10 \; \mbox{ps}
                    \end{array} \right.
             \hspace{0.5cm} \mbox{for} 
                  \left. \begin{array}{ll}
\bar{m}_c(\bar{m}_b), &\bar{m}_b(\bar{m}_b)
                    \end{array} \right. \; .
\label{taubm2}
\end{eqnarray}
Going to next-to-leading logarithmic accuracy we have to use the Wilson coefficients
of the effective Hamiltonian to NLO accuracy and we have to determine one-loop QCD corrections
within the effective theory.
These NLO-QCD corrections turned out to be very important for the inclusive
$b$-quark decays. For massless final state quarks
the calculation was done in 1991 \cite{Altarelli:1991dx}: 
\begin{equation}
c_{3,b} = c_{3,b}^{\rm LO-QCD} + 8\frac{\alpha_s}{4 \pi}
\left[
\left(\frac{25}{4} - \pi^2 \right)
+
2 \left(C_1^2 + C_2^2 \right) \left(\frac{31}{4} - \pi^2 \right)
- \frac{4}{3} C_1 C_2         \left(\frac{7}{4} + \pi^2 \right)
\right] \; .
\label{c3NLO}
\end{equation}
The first QCD corrections in Eq.(\ref{c3NLO}) stems from semi-leptonic decays.
It can be guessed from the correction to the muon decay in Eq.(\ref{murad}) 
by decomposing the factor 8 in Eq.(\ref{c3NLO}) as $8 = 3 \cdot C_F \cdot 2$:
$3$ comes from the three leptons $e^-, \mu^-, \tau^-$, $C_F$ is a QCD colour factor
and $2$ belongs to the correction in Eq.(\ref{murad}).
The second and the third term in Eq.(\ref{c3NLO}) stem from non-leptonic decays.
\\
It turned out, however, that effects of the charm quark mass are crucial, 
see, e.g., the estimate in \cite{Voloshin:1994sn}. 
NLO-QCD corrections with full mass dependence 
were determined 
for $b \to c l^- \bar{\nu}$ already in 1983 \cite{Hokim:1983yt},
for $b \to c \bar{u} d   $ in 1994 \cite{Bagan:1994zd},
for $b \to c \bar{c} s   $ in 1995 \cite{Bagan:1995yf},
for $b \to $ no charm      in 1997 \cite{Lenz:1997aa}
and 
for $b \to s g $           in 2000 \cite{Greub:2000sy,Greub:2000an}.
Since there were several misprints in \cite{Bagan:1995yf}- leading to IR divergent expressions -, 
the corresponding calculation was redone in \cite{Krinner:2013cja} and the numerical result was 
updated.\footnote{The authors of \cite{Bagan:1995yf} left particle physics and it was not possible
to obtain the correct analytic expressions. The numerical results in \cite{Bagan:1995yf} were, 
however, correct.}
With the results in \cite{Krinner:2013cja}
we predict (using $\bar{m}_c(\bar{m}_b)$ and  $\bar{m}_b(\bar{m}_b)$)
\begin{equation}
        c_{3,b}               =  
\left \{ 
         \begin{array}{ll}
           9            & (m_c = 0 = \alpha_s)
           \\
           5.29 \pm 0.35  & ({\rm LO-QCD}) %5.28619\pm 0.35453
           \\
           6.88 \pm 0.74  & ({\rm NLO-QCD})
         \end{array}
          \; .
        \right.
\label{c3numfinal}
        \end{equation}
        Comparing this result with Eq.(\ref{c3num}) one finds a huge phase space suppression,
        which reduces the value of $C_{3,b}$ from 9 in the mass less case to about 4.7 when 
        including charm quark mass effect. Switching on in addition QCD effects $c_{3,b}$ is 
        enhanced back to a value of about 6.9.
        The LO $b \to c$ transitions contribute about $70\%$ to this value, 
        the full NLO-QCD corrections
        about $24 \%$ and the $b \to u$ and penguin contributions about $6 \%$ 
        \cite{Krinner:2013cja}.
\\
For the total lifetime we predict thus
\begin{eqnarray}
\tau_b = (1.65 \pm 0.24)  \; \mbox{ps} \; ,
% 1.6465 \pm 0.240863
\label{taubfull}
\end{eqnarray}
which is our final number for the lifetime of a free $b$-quark.
This number is now very close to the 
experimental numbers in Eq.(\ref{tauBexp}), unfortunately
the uncertainty is still quite large. To reduce this, a calculation at the NNL order 
would be necessary. Such an endeavour seems to be doable nowadays. The dominant Wilson coefficients
$C_1$ and $C_2$ are known at NNLO accuracy \cite{Gorbahn:2004my} and
the two loop corrections in the effective theory have been determined e.g. in
\cite{vanRitbergen:1999gs,Melnikov:2008qs,Pak:2008qt,Pak:2008cp,Bonciani:2008wf,Biswas:2009rb}
for semi-leptonic decays and partly in
\cite{Czarnecki:2005vr}
for non-leptonic decays.
\\
It is amusing to note, that a naive treatment with vanishing charm quark masses and 
neglecting the sizable QCD-effects, 
see Eq.(\ref{taubm0}), yields by accident a similar result as in Eq.(\ref{taubfull}).
The same holds also for the semi leptonic branching ratio, where a naive treatment
($m_c = 0 = \alpha_s$) gives
\begin{equation}
B_{sl} = \frac{\Gamma (b \to c e^- \bar{\nu}_e )}{\Gamma_{\rm tot}}
= \frac{1}{9} = 11.1 \% \; ,
\end{equation}
while the full treatment (following \cite{Krinner:2013cja}) gives
\begin{equation}
B_{sl} = (11.6 \pm 0.8) \% \; .
%0.117276 \pm 0.00809575
\label{Bsltheo}
\end{equation}
This number agrees well with recent measurements \cite{PDG,Oswald:2012yx}
\begin{eqnarray}
B_{sl}(B_d) & = & (10.33 \pm 0.28) \% \; ,
\nonumber
\\
B_{sl}(B^+) & = & (10.99 \pm 0.28) \% \; ,
\label{Bslexp}
\\
B_{sl}(B_s) & = & (10.61 \pm 0.89) \% \; .
\nonumber
\end{eqnarray}

\subsubsection{The HQE}
Now we are ready to derive the heavy quark expansion for inclusive 
decays.\footnote{We delay almost all referencing
related to the creation of the HQE to Section \ref{lifetime_sec3}.}
The decay rate of the transition of a B-meson to an inclusive final state $X$
can be expressed as a phase space integral over the square of
the matrix element of the effective Hamiltonian sandwiched between
the initial $B$-meson\footnote{The replacements one has to do when considering
a $D$-meson decay are either trivial or we explicitly comment on them.}
state and the final state $X$.
Summing over all final states $X$ with the same quark quantum numbers 
we obtain
 \begin{equation}
 \Gamma ( B \to X) = \frac{1}{2 m_B} \sum \limits_{X} \int_{\rm PS} (2 \pi)^4 \delta^{(4)}
 (p_B - p_X) | \langle X | {\cal H}_{eff} | B \rangle |^2 
\, .
\label{total}
 \end{equation}
If we consider, e.g., a decay into three particles, i.e. $B \to 1+2+3$, then the phase space
integral reads
\begin{equation}
\int_{\rm PS} = \prod \limits_{i=1}^3 \left[ \frac{d^3 p_i}{(2 \pi)^3 2 E_i} \right]
\end{equation} 
and $p_X = p_1+p_2+p_3$.
With the help of the optical theorem the total decay rate in Eq.(\ref{total})  can be rewritten as
 \begin{equation}
 \Gamma(B \to X) = \frac{1}{2 m_B} \langle B |{\cal T} | B \rangle
\, ,
 \end{equation}
with the transition operator
 \begin{equation}
 {\cal T} = \mbox{Im} \; i \int d^4x 
 T \left[ {\cal H}_{eff}(x) {\cal H}_{eff} (0) \right] 
\, ,
\label{trans}
 \end{equation}
consisting of a non-local double insertion of the effective Hamiltonian.
\\
A second operator-product-expansion, exploiting the large value of the $b$-quark mass $m_b$, 
yields for
 $ {\cal{T}}$
  \begin{equation}
 {\cal T} = \frac{G_F^2 m_b^5}{192 \pi^3} |V_{cb}|^2
\left[{ c_{3,b}} { \bar{b}b} + 
              \frac{c_{5,b}}{m_b^2} { \bar{b}g_s \sigma_{\mu \nu} G^{\mu \nu} b} +
            2 \frac{c_{6,b}}{m_b^3} { (\bar{b}q)_{\Gamma} (\bar{q}b)_{\Gamma}} 
            +...  \right]
\, 
 \end{equation}
and thus for the decay rate
  \begin{equation}
\boxed{
\Gamma =   \frac{G_F^2 m_b^5}{192 \pi^3} |V_{cb}|^2
\left[
  c_{3,b}             \frac{\langle B | \bar{b}b| B \rangle}{2 M_B} 
+ 
\frac{c_{5,b}}{m_b^2} \frac{\langle B | \bar{b}g_s \sigma_{\mu \nu} G^{\mu \nu} b| B \rangle}{2 M_B}
+ 
\frac{c_{6,b}}{m_b^3} \frac{\langle B | (\bar{b}q)_{\Gamma} (\bar{q}b)_{\Gamma}| B \rangle}{M_B}
            +... 
\right]
\label{HQE}
}
\, .
 \end{equation}
The individual contributions in Eq.(\ref{HQE}) have the following origin and interpretation:
\\
\\
\underline{\bf Leading term in Eq.(\ref{HQE}):}
\\
\\
To get the first term we contracted all quark lines, 
        except the beauty-quark lines, in the
        product of the two effective Hamiltonians. This leads to the following 
        two-loop diagram on the 
        l.h.s., where the circles with the crosses denote the $\Delta B = 1$-operators 
        from the effective 
        Hamiltonian.

\vspace{0.3cm}

\centerline{\includegraphics[width=  \textwidth, angle=0]{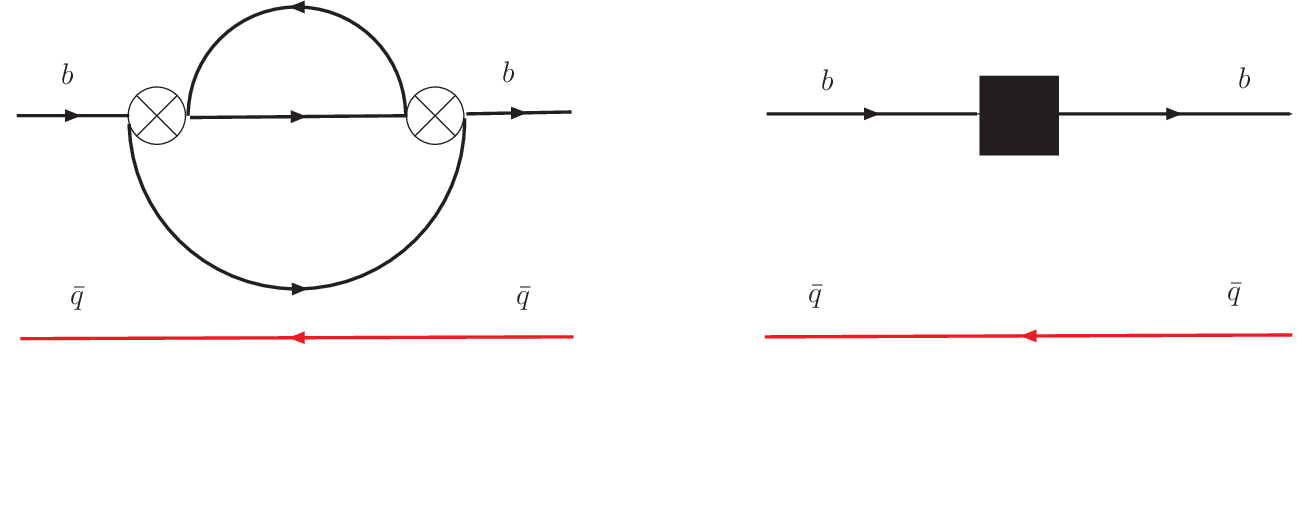}}

\vspace{-0.8cm}

\noindent
        Performing the loop 
        integrations in this diagram we get the Wilson coefficient $c_{3,b}$
        that contains all the loop functions and the dimension-three
        operator { $\bar{b}b$}, which is denoted by the black square in the diagram on the r.h.s. . 
        This has been done already in Eq.(\ref{c3LOQCD}),  Eq.(\ref{c3NLO}) and Eq.(\ref{c3numfinal}).  
        \\
        A crucial finding for the HQE was the fact, that the matrix element of the 
        dimension-three operator $\bar{b} b$ can 
        also be expanded in the
        inverse of the $b$-quark mass. According to the  
        Heavy Quark Effective Theory (HQET) we get\footnote{We use here the conventional 
        relativistic normalisation $\langle B | B \rangle =2 E V $, where $E$ denotes the 
        energy of the meson and $V$ the space volume.
        In the original literature sometimes
        different normalisations have been used, which can lead to confusion.}
        \begin{equation}
         \frac{\langle B | \bar{b}b | B \rangle}{2 M_B} = 
          1 - \frac{ \mu_\pi^2 -  \mu_G^2}{2m_b^2} 
             + {\cal O} \left(\frac{1}{m_b^3} \right)\; ,
        \label{bbarbME}
        \end{equation}
        with the matrix element of the kinetic operator $ \mu_\pi^2$ and the matrix element of the
        chromo-magnetic operator $ \mu_G^2$, defined in the $B$-rest frame 
        as\footnote{We use here $\sigma_{\mu \nu} =
\frac{i}{2} [\gamma_\mu, \gamma_\nu]$. In the original literature sometimes the notation $
         i \sigma G := i \gamma_\mu \gamma_\nu G^{\mu \nu}$ was used, which differs by a factor of 
         $i$ from our definition of $\sigma$.}
        \begin{eqnarray}
         \mu_\pi^2 & = &  \frac{\langle B | \bar{b} ( i \vec{D})^2 b | B \rangle}{2 M_B} 
         + {\cal O} \left(\frac{1}{m_b} \right)\; ,
         \\
         \mu_G^2  & = &  
         \frac{\langle B | \bar{b} \frac{g_s}{2} \sigma_{\mu \nu} G^{\mu \nu} b | B \rangle}{2 M_B} 
        + {\cal O} \left(\frac{1}{m_b} \right)\; .
        \end{eqnarray}
       With the above definitions for the non-perturbative matrix-elements the expression for the total 
       decay rate in Eq.(\ref{HQE}) becomes
       \begin{eqnarray}
       \Gamma = \frac{G_F^2 m_b^5}{192 \pi^3} V_{cb}^2
                && \left\{ 
                   c_{3,b}      \left[  1 - \frac{ \mu_\pi^2 -  \mu_G^2}{2m_b^2} 
                   + {\cal O} \left(\frac{1}{m_b^3} \right) \right]
                   \right.
         \nonumber \\
        && \left.
+ 2 c_{5,b}
\left[ \frac{\mu_G^2}{m_b^2}  + {\cal O} \left(\frac{1}{m_b^3} \right) \right]
+
\frac{c_{6,b}}{m_b^3} \frac{\langle B | (\bar{b}q)_{\Gamma} (\bar{q}b)_{\Gamma}| B \rangle}{M_B}
            +... 
\right\}
\, .
\label{HQE2}
 \end{eqnarray}
The leading term in Eq.(\ref{HQE2}) describes simply the decay of a free quark. 
Since here  the spectator-quark (red) is not involved in the decay process at all, 
this contribution will be the same for all different b-hadrons, thus predicting the same 
lifetime for all $b$-hadrons.
\\
The first corrections are already suppressed by two powers of the heavy $b$-quark
mass - we have no corrections of order $1/m_b$!
This non-trivial result explains, why our description in terms of the free b-quark
decay was so close to the experimental values of the lifetimes of $B$-mesons.
\\
In the case of $D$-mesons the expansion parameter $1/m_c$ is not small
and the higher order terms of the HQE will lead to sizable corrections.
The leading term $c_{3,c}$ for charm-quark decays gives at the scale $\mu = M_W$ for
vanishing quark mass $c_{3,c} = 5$. At the scale $\mu = \bar{m}_c (\bar{m}_c)$ and
realistic values of final states masses we get
\begin{eqnarray}
c_{3,c} & = &
         \left \{ 
         \begin{array}{ll}
           5               & (m_s = 0 = \alpha_s)
           \\
            6.29 \pm 0.72   & ({\rm LO-QCD})
           \\
           11.61 \pm 1.55  & ({\rm NLO-QCD})
         \end{array}
          \; .
        \right.
\end{eqnarray}
Here we have a large QCD enhancement of more than a factor of two, while phase space effects seem
to be negligible.
\\
The  $1/m_b^2$-corrections  in Eq.(\ref{HQE2}) have two sources: first the 
expansion in Eq.(\ref{bbarbME}) and the second one 
- denoted by the term proportional to $c_{5,b}$ - 
will be discussed below.
\\
Concerning the different $1/m_b^3$-corrections, indicated in  Eq.(\ref{HQE2}),
we will see that the first two terms of the expansion in Eq.(\ref{HQE})
are triggered by a two-loop diagram, while the third term is given by a one-loop
diagram. This will motivate, why the $1/m_b^3$-corrections proportional to 
$c_{3,b}$ and $c_{5,b}$ can be neglected in comparison to the  $1/m_b^3$-corrections 
proportional to $c_{6,b}$; the former ones 
will, however, be important
for precision determination of semi-leptonic decay rates\footnote{Kolya made substantial contributions
to these higher order terms, which will be discussed somewhere else in these book. For our purpose
of investigating lifetimes they can, however, be safely neglected, because there the hadronic 
uncertainties are still considerably larger.}.  
\\
\\
\\
\underline{\bf Second term in Eq.(\ref{HQE}):}
\\
\\
To get the second term in Eq.(\ref{HQE}) we couple in addition a gluon to the vacuum.
This is denoted by the diagram below, where a gluon is emitted from one of the internal 
quarks of the two-loop diagram. Doing so, we obtain the so-called
chromo-magnetic operator { $\bar{b}g_s \sigma_{\mu \nu} G^{\mu \nu} b$}, which already appeared
in the expansion in Eq.(\ref{bbarbME}). 

\vspace{0.3cm}

\centerline{\includegraphics[width= \textwidth, angle =0]{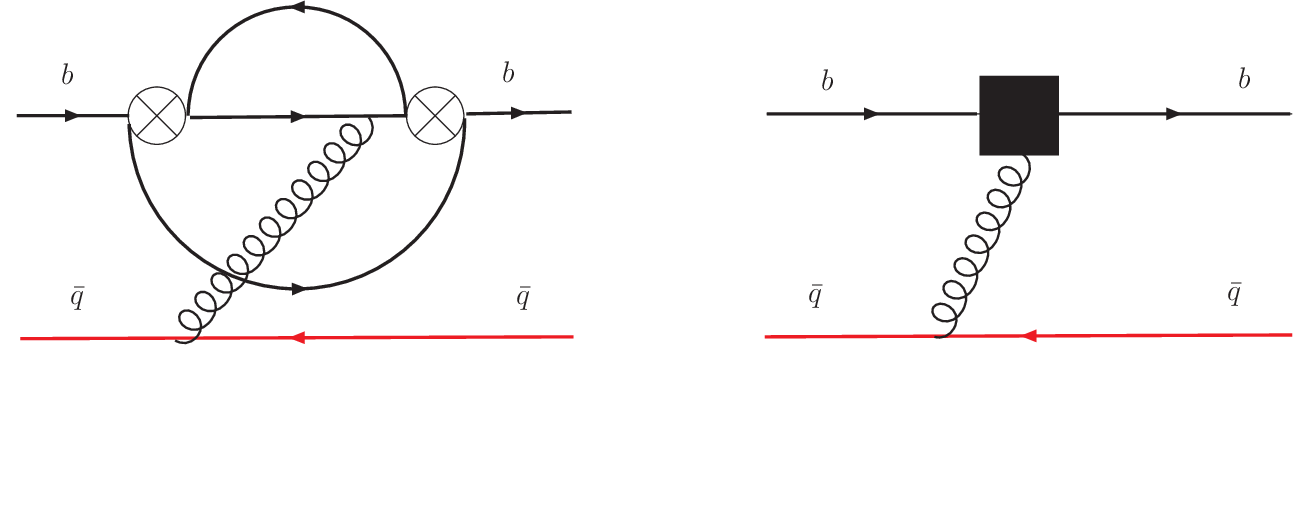}}

\vspace{-1cm}
 
\noindent
        Since this
        operator is of  dimension five, the corresponding 
        contribution is - as seen before -
        suppressed by two powers of the heavy quark mass, compared to the 
        leading term.
        The corresponding Wilson coefficient $c_{5,b}$ reads, e.g., for the semi-leptonic decay 
        $b \to c e^- \bar{\nu}_e $\footnote{The result in Eq.(94) of the review
        \cite{Neubert:1997gu} has an additional factor $6$ in $c_5^{ce\bar{\nu}_e}$.} 
        and  the non-leptonic decays $b \to c \bar{u} d$ and  $b \to c \bar{c} s$
        \begin{eqnarray}
         c_{5,b}^{ce\bar{\nu}_e} & = & - \left(1 - z \right)^4  \left[ 1+ \frac{\alpha_s}{4 \pi} \dots
        \right]\; ,
         \label{c5sl}
         \\
         c_{5,b}^{ c \bar{u} d} & = & -  \left|V_{ud}\right|^2  
                    \left(1 - z \right)^3
        \left[ {\cal N}_a (\mu) \left(1 -z \right)
               + 8 C_1 C_2 + \frac{\alpha_s}{4 \pi} \dots
        \right] \; ,
         \label{c5nl}
         \\
         c_{5,b}^{ c \bar{c} s} & = & -  \left| V_{cs}\right|^2  
         \left\{ {\cal N}_a (\mu) 
         \left[ \sqrt{1 - 4z} (1 - 2 z) (1 - 4 z - 6 z^2)+ 24 z^4
                \log \left(\frac{1 + \sqrt{1 - 4z}}{1 - \sqrt{1 - 4z}}\right)
         \right]
%       \frac{1 - 10z + 26z^2 + 4z^3 - 48z^4 + 24 z^4 \sqrt{1 - 4z}
%             \log \left[\frac{1 + \sqrt{1 - 4z}}{1 - \sqrt{1 - 4z}}\right]}{\sqrt{1 - 4z}}
             \right.
          \nonumber 
          \\
          && \left. \hspace{0.1cm}
               + 8 C_1 C_2 \left[\sqrt{1 - 4z} \left(1+\frac{z}{2} + 3 z^2 \right) - 3 z (1-2 z^2)
           \log \left(\frac{1 + \sqrt{1 - 4z}}{1 - \sqrt{1 - 4z}}\right) \right]
           + \frac{\alpha_s}{4 \pi} \dots
        \right\} \; ,
        \nonumber
        \\       
  \label{c5nl2}
        \end{eqnarray}
with the quark mass ratio $z = (m_c/m_b)^2$. 
For vanishing charm-quark masses and $V_{ud} \approx 1$ 
we get $c_{5,b}^{ c \bar{u} d}  = -3$ at the scale $\mu = M_W$, which reduces in LO-QCD
to about $-1.2$ at the scale $\mu = m_b$.
\\
For the total decay rate we have to sum up all possible quark level-decays
        \begin{equation}
        c_{5,b} = c_{5,b}^{c e \bar{\nu}_e} +  
                  c_{5,b}^{c \mu \bar{\nu}_\mu} +  
                  c_{5,b}^{c \tau \bar{\nu}_\tau} +   
                  c_{5,b}^{ c \bar{u} d} +  
                  c_{5,b}^{ c \bar{c} s} + \dots \; . 
        \end{equation}
Neglecting penguin contributions we get numerically 
         \begin{equation}
         c_{5,b} = \left\{
             \begin{array}{ll}
             \approx -  9  & (m_c = 0 = \alpha_s)
             \\
             -3.8 \pm  0.3 & ( \bar{m}_c (\bar{m}_b) \; , \alpha_s(m_b) )
             \end{array}
            \right.             \; ,
        \end{equation}
For $c_{5,b}$ both QCD effects as well as phase space effects are quite pronounced.
The overall coefficient of the matrix element
of the chromo-magnetic operator $\mu_G^2$ normalised to $2m_b^2$ in Eq.(\ref{HQE2}) is given by 
$c_{3,b} + 4 c_{5,b}$, which is sometimes
denoted as $c_{G,b}$. For semi-leptonic decays like  $b \to c e^- \bar{\nu}_e $, it 
reads\footnote{We differ here slightly from Eq.(7) of \cite{Luke:1993za}, who have 
a different sign in the coefficients of $z^2$ and $z^3$. We agree, however, with the corresponding 
result in \cite{Falk:1994gw}.}
        \begin{eqnarray}
         c_{G,b}^{ce\bar{\nu}_e} =  c_{3,b}^{ce\bar{\nu}_e} + 4 c_{5,b}^{ce\bar{\nu}_e}
         & = & (-3) \left[1 - \frac83 z + 8 z^2 - 8 z^3 + \frac53 z^4 + 4 z^2 \ln (z)
        \right] 
        \; .
         \end{eqnarray}
For the sum of all inclusive decays we get
         \begin{equation}
         c_{G,b} = \left\{
             \begin{array}{ll}
             -27 = -3 c_3           & ( m_c = 0 = \alpha_s)
             \\
             -7.9 \approx - 1.1 c_3 & (\bar{m_c} (\bar{m_c}) \; , \alpha_s(m_b) ) 
             \end{array}
            \right.             \; ,
         \label{cG}
        \end{equation}
leading to the following form of the total decay rate
\begin{equation}
\boxed{
\Gamma = \frac{G_F^2 m_b^5}{192 \pi^3} V_{cb}^2
               \left[ 
                   c_{3,b}   
                -  c_{3,b}  \frac{\mu_\pi^2}{2m_b^2} 
                +  c_{G,b}  \frac{\mu_G^2  }{2m_b^2} 
                +  \frac{c_{6,b}}{m_b^3} \frac{\langle B | (\bar{b}q)_{\Gamma} (\bar{q}b)_{\Gamma}| B \rangle}{M_B}
            +... 
\right]
}
\label{HQE3} \; .
 \end{equation}
Both $1/m_b^2$-corrections are reducing the decay rate and their overall coefficients are of 
similar size as $c_{3,b}$.
To estimate more precisely the numerical effect of the $1/m_b^2$ corrections, 
we still need the values of 
$\mu_\pi^2$ and $\mu_G^2$.
 Current values \cite{Gambino:2013rza,Uraltsev:2001ih} of these parameters read for the case
        of $B_d$ and $B^+$-mesons
        \begin{eqnarray}
         \mu_\pi^2 (B)& = &   (0.414 \pm 0.078) \; {\rm GeV}^2 
         \; ,
         \label{mupi}
         \\
         \mu_G^2 (B) & \approx & \frac{3}{4} \left( M_{B^*}^2 - M_B^2 \right) 
                             \approx (0.35 \pm 0.07) \; {\rm GeV}^2  
        \; .
         \label{muG}
        \end{eqnarray}
        For $B_s$-mesons only small differences compared to  $B_d$ and $B^+$-mesons 
        are predicted \cite{Bigi:2011gf}
        \begin{eqnarray}
         \mu_\pi^2 (B_s)-\mu_\pi^2 (B_d)& \approx &   (0.08 \dots 0.10) \; {\rm GeV}^2 
         \; ,
         \\
         \frac{\mu_G^2 (B_s)}{\mu_G^2 (B_d)} & \approx & 1.07 \pm 0.03
        \; ,
        \end{eqnarray}
        while sizable differences  are expected \cite{Bigi:2011gf} for $\Lambda_b$-baryons.
        \begin{eqnarray}
         \mu_\pi^2 (\Lambda_b)-\mu_\pi^2 (B_d)& \approx &   (0.1 \pm 0.1) \; {\rm GeV}^2 
         \; ,
         \\
         \mu_G^2 (\Lambda_b) & = & 0
        \; .
        \end{eqnarray}
Inserting these values in Eq.(\ref{HQE3})
we find that the $1/m_b^2$-corrections are decreasing the decay rate slightly
($m_b = \bar{m}_b(\bar{m}_b) = 4.248$ GeV): 
\begin{equation}
\begin{array}{|c||c|c|c|c|}
\hline
       &   B_d  & B^+  & B_s  & \Lambda_d
\\
\hline
\hline
-\frac{\mu_\pi^2}{2 m_b^2} & -0.011 & -0.011 & -0.014 & -0.014
\\
\hline
\frac{c_{G,b}}{c_{3,b}} \frac{\mu_\pi^2}{2 m_b^2}  & -0.011 & -0.011 & -0.011 & 0.00
\\
\hline
\end{array}
\end{equation}
The kinetic and the chromo-magnetic operator each reduce the decay rate by about $1\%$,
except for the case of the $\Lambda_b$-baryon, where the chromo-magnetic operator 
vanishes.
The $1/m_b^2$-corrections exhibit now also a small sensitivity to the spectator-quark. 
Different values for the lifetimes of $b$-hadrons can arise due to 
different values of the non-perturbative parameters $\mu_G^2$ and $\mu_\pi^2$, 
the corresponding numerical effect will, however, be small. 
\begin{equation}
\begin{array}{|c||c|c|c|c|}
\hline
  X:                                                       & B^+     & B_s  & \Lambda_d
\\
\hline
\hline
\frac{\mu_\pi^2(X) - \mu_\pi^2(B_d)}{2 m_b^2}              &  0.000 \pm 0.000 &  0.002 \pm 0.000 &  0.003 \pm 0.003
\\
\hline
\frac{c_{G,b}}{c_{3,b}} \frac{\mu_G^2(X) - \mu_G^2(B_d)}{2 m_b^2}  &  0.000 \pm 0.000 & 0.000...0.001 &  -0.011 \pm 0.003
\\
\hline
\end{array}
\label{delta1mb2}
\end{equation}
Thus we find that the $1/m_b^2$-corrections give no difference in the lifetimes 
of $B^+$- and $B_d$-mesons, 
they enhance the $B_s$-lifetime
by about $3$ per mille, compared to the $B_d$-lifetime and they reduce the $\Lambda_b$-lifetime
by about $1 \%$ compared to the $B_d$-lifetime.
\\
To get an idea of the size of these corrections in the charm-system, we first 
investigate the Wilson coefficient $c_5$. 
         \begin{equation}
         c_{5,c} = \left\{
             \begin{array}{ll}
             \approx -  5  & (m_c = 0 = \alpha_s)
             \\
             -1.7 \pm  0.3 & ( \bar{m_c} (\bar{m_c}) \; , \alpha_s(m_b) )
             \end{array}
            \right.             \; ,
        \end{equation}
At the scale $\mu = m_c$ the non-leptonic contribution to $c_5$ is getting smaller than in the
bottom case and it even changes sign. 
For the coefficient $c_G$ we find
         \begin{equation}
         c_{G,c} = \left\{
             \begin{array}{ll}
             \approx -  15 = -3 \, c_{3,c} & (m_c = 0 = \alpha_s)
             \\
             4.15 \pm  1.48 = ( 0.37 \pm 0.13) \, c_{3,c} & ( \bar{m_c} (\bar{m_c}) \; , \alpha_s(m_b) )
             \end{array}
            \right.             \; .
        \end{equation}
We see for that for the charm case the overall coefficient of the chromo-magnetic 
operator has now a positive sign and the relative size is less than in the bottom case.
For $D^0$- and  $D^+$-mesons
the value of the chromo-magnetic operator reads
\begin{equation}
\mu_G^2 (D)  \approx  \frac{3}{4} \left( M_{D^*}^2 - M_D^2 \right) 
                             \approx 0.41 \; {\rm GeV}^2  
        \; ,
\label{muGD}
\end{equation}
which is of similar size as in the $B$-system. Normalising this value to the charm quark mass
$m_c = \bar{m}_c(\bar{m}_c) = 1.277$ GeV, we get however a bigger contribution compared to the 
bottom case and also a different sign.
\begin{equation}
c_{G,c} \frac{\mu_G^2 (D)}{2 m_c^2} \approx + 0.05 \, c_{3,c} \; .
\end{equation}
Now the second order corrections are non-negligible, with a typical size of about + $5 \%$ 
of the total decay rate. Concerning lifetime differences of $D$-mesons, we find no visible 
effect due to the 
chromo-magnetic operator \cite{Lenz:2013aua}
\begin{eqnarray}
\frac{\mu_G^2 (D^+)}{\mu_G^2 (D^0)} & \approx & 0.993 \; ,
\\
\frac{\mu_G^2 (D^+_s)}{\mu_G^2 (D^0)} & \approx & 1.012 \pm 0.003 \; .
\end{eqnarray}
For the kinetic operator a sizable SU(3) flavour breaking was found by Bigi, Mannel and 
Uraltsev \cite{Bigi:2011gf}
\begin{eqnarray}
\mu_\pi^2 (D^+_s) - \mu_\pi^2 (D^0) & \approx & 0.1  \;  \mbox{GeV}^2 \; ,
\end{eqnarray}
leading to an reduction of the $D^+_s$-lifetime of the order of $3 \%$ compared to the $D^0$-lifetime 
\begin{eqnarray}
\frac{\mu_\pi^2 (D^+_s) - \mu_\pi^2 (D^0)}{2 m_c^2} & \approx & 0.03 \; .
\end{eqnarray}
%\\
%\\
\underline{\bf Third term in Eq.(\ref{HQE}):}
\\
\\
The next term is obtained by only contracting two quark lines in the product
of the two effective Hamiltonian in Eq.(\ref{trans}). The $b$-quark and the
spectator quark of the considered hadron are not contracted. For $B_d$-mesons ($q=d$)
and  $B_s$-mesons ($q=s$) we get the following so-called 
{\it weak annihilation} diagram.

\centerline{\includegraphics[width=\textwidth, angle =0]{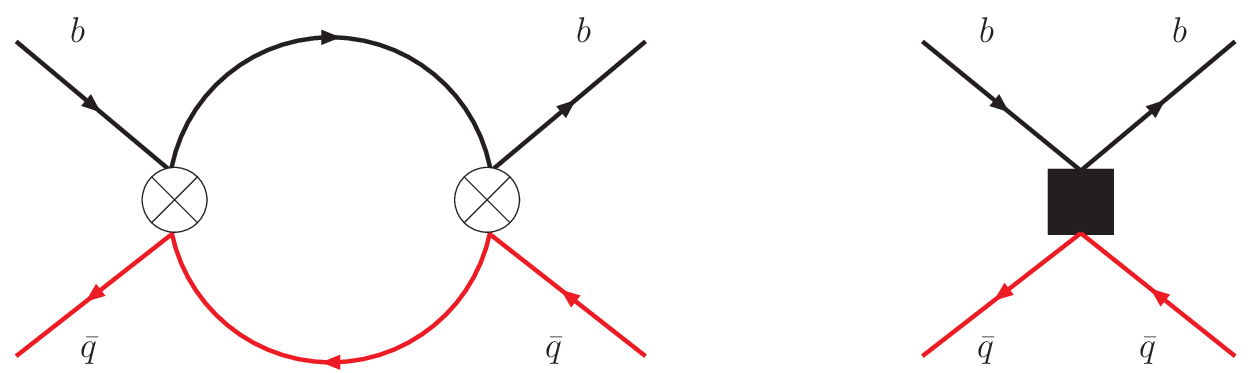}}

%\centerline{\includegraphics[width=\textwidth, angle =0]{Effective.eps}}

\noindent
Performing the loop integration on the diagram on the l.h.s. we get the Wilson coefficient
$c_6$ and dimension six four-quark operators
$(\bar{b}q)_{\Gamma} (\bar{q}b)_{\Gamma}$, with Dirac structures $\Gamma$.
The corresponding matrix elements of these $\Delta B = 0$ operators are typically written as
\begin{equation}
\langle B | (\bar{b}q)_{\Gamma} (\bar{q}b)_{\Gamma} | B \rangle =  
c_\Gamma f_B^2 M_B  B_\Gamma \; ,
\end{equation}
with the bag parameter $B_\Gamma$, the decay constant $f_B$ and a numerical factor
$c_\Gamma$ that contains some colour factors and sometimes also ratios of masses.
\\
For the case of the $B^+$-meson we get a similar diagram, with the only difference that
now the external spectator-quark lines are crossed, this is the so-called 
{\it Pauli interference} diagram.

%\centerline{\includegraphics[width=\textwidth, angle =0]{dim6matching.eps}}

\centerline{\includegraphics[width=0.6\textwidth, angle =0]{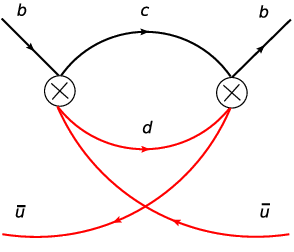}}

There are two very interesting things to note. First this is now a 
one-loop diagram. Although being suppressed by three powers of the
$b$-quark mass it is enhanced by a phase space factor of $16 \pi^2$ compared to the leading
two-loop diagrams. Second, now we are really sensitive to the 
flavour of the spectator-quark, because in principle, each different spectator 
quark gives a different contribution\footnote{This difference is, however, negligible, 
if one considers, e.g., $B_s$ vs. $B_d$.}.
These observations are responsible for the fact that lifetime 
differences in the system of heavy hadrons are almost entirely due to
the contribution of weak annihilation and Pauli interference diagrams.
\\
In the case of the $B_d$ meson four different four-quark operators
arise
\begin{eqnarray}
Q^q   = \bar{b} \gamma_\mu (1-\gamma_5) q     \times \bar{q} \gamma^\mu (1-\gamma_5) b  ,
&&
Q^q_S = \bar{b} (1-\gamma_5) q                \times \bar{q} (1-\gamma_5) b  ,
\label{operator1}
\\
T^q   = \bar{b} \gamma_\mu (1-\gamma_5) T^a q \times \bar{q} \gamma^\mu (1-\gamma_5) T^a b  ,
&&
T^q_S = \bar{b} (1-\gamma_5) T^a q            \times \bar{q} (1-\gamma_5)T^a b  ,
\; ,
\label{operator2}
\end{eqnarray}
with $q=d$ for the case of $B_d$-mesons. $Q$ denotes colour singlet operators and 
$T$ colour octet operators. For historic reasons the matrix elements of these operator are 
typically expressed as
\begin{eqnarray}
\frac{\langle B_d |Q^d   | B_d \rangle}{M_{B_d}} =  f_B^2 B_1 M_{B_d} \; ,
&&
\frac{\langle B_d |Q^d_S | B_d \rangle}{M_{B_d}} = f_B^2 B_2 M_{B_d} \; ,
\label{Bag1}
\\
\frac{\langle B_d |T^d   | B_d \rangle}{M_{B_d}} =  f_B^2 \epsilon_1 M_{B_d} \; ,
&&
\frac{\langle B_d |T^d_S | B_d \rangle}{M_{B_d}} =  f_B^2 \epsilon_2 M_{B_d} \; .
\label{Bag2}
\end{eqnarray}
The bag parameters $B_{1,2}$ are expected to be of order one in vacuum insertion approximation,
while the $\epsilon_{1,2}$ vanish in that limit. We will discuss below several estimates
of $B_i$ and $\epsilon_i$.
Decay constants can be determined with lattice-QCD, see, e.g., the reviews of 
FLAG \cite{Colangelo:2010et} or with QCD sum rules, see, e.g., the recent determination 
in \cite{Gelhausen:2013wia}.
Later on, we will see, however, that the 
Wilson coefficients of $B_1$ and $B_2$ are affected by sizable numerical cancellations, enhancing
hence the relative contribution of the colour suppressed $\epsilon_1$ and $\epsilon_2$.
The corresponding Wilson coefficients of the four operators can be written as
\begin{eqnarray}
 c_6^{Q^d}     =  16 \pi^2 \left[ \left| V_{ud}\right|^2 F^u + \left| V_{cd}\right|^2 F^c  \right]  ,
&&
 c_6^{Q^d_S}   =  16 \pi^2 \left[ \left| V_{ud}\right|^2 F^u_S +\left| V_{cd}\right|^2 F^c_S\right]  ,
\\
 c_6^{T^d}     =  16 \pi^2 \left[ \left| V_{ud}\right|^2 G^u   +\left| V_{cd}\right|^2 G^c\right]  ,
&&
 c_6^{T^d_S}   =  16 \pi^2 \left[ \left| V_{ud}\right|^2 G^u_S +\left| V_{cd}\right|^2 G^c_S\right]  .
\end{eqnarray}  
$F^q$ describes an internal $c\bar{q}$ loop in the above weak annihilation diagram.
The functions $F$ and $G$ are typically split up in contributions proportional to $C_2^2$, $C_1 C_2$
and $C_1^2$.
\begin{eqnarray}
F^u & = & C_1^2 F_{11}^u + C_1 C_2 F_{12}^u +   C_2^2 F_{22}^u \; ,
\\
F^u_S & = & \dots \; .
\end{eqnarray}
Next, each of the $F^q_{ij}$ can be expanded in the strong coupling
\begin{eqnarray}
F^u_{ij} & = & F^{u,(0)}_{ij} + \frac{\alpha_s}{4 \pi} F^{u,(1)}_{ij} + \dots \; ,
\\
F^u_{S,ij} & = & \dots \; .
\end{eqnarray}
As an example we give the following LO results
\begin{eqnarray}
F^{u,(0)}_{11}   = -3 (1-z)^2 \left( 1 + \frac{z}{2} \right) \; ,
&&
F^{u,(0)}_{S,11} =  3 (1-z)^2 \left( 1 + 2 z \right) \; ,
\\
F^{u,(0)}_{12}   = -2 (1-z)^2 \left( 1 + \frac{z}{2} \right) \; ,
&&
F^{u,(0)}_{S,12} =  2 (1-z)^2 \left( 1 + 2 z \right) \; ,
\\
F^{u,(0)}_{22}   = -\frac13 (1-z)^2 \left( 1 + \frac{z}{2} \right) \; ,
&&
F^{u,(0)}_{S,22} =  \frac13 (1-z)^2 \left( 1 + 2 z \right) \; ,
\\
G^{u,(0)}_{22}   = -2 (1-z)^2 \left( 1 + \frac{z}{2} \right) \; ,
&&
G^{u,(0)}_{S,22} =  2 (1-z)^2 \left( 1 + 2 z \right) \; ,
\end{eqnarray}
with $z = m_c^2/m_b^2$.
\\
Putting everything together we arrive at the
following expression for the decay rate of a $B_d$-meson
\begin{eqnarray}
\Gamma_{B_d} &= &\frac{G_F^2 m_b^5}{192 \pi^3} V_{cb}^2
               \left[ 
                   c_3   
                -  c_3  \frac{\mu_\pi^2}{2m_b^2} 
                +  c_G  \frac{\mu_G^2  }{2m_b^2} 
                + \frac{16 \pi^2 f_B^2 M_{B_d}}{m_b^3} \tilde{c}_6^{B_d}
                   + {\cal O} \left( \frac{1}{m_b^3},  \frac{16 \pi^2}{m_b^4}\right) \right]
\nonumber \\
 & \approx &\frac{G_F^2 m_b^5}{192 \pi^3} V_{cb}^2
               \left[ 
                   c_3   
                -  0.01 c_3  
                -  0.01 c_3 
                + \frac{16 \pi^2 f_B^2 M_{B_d}}{m_b^3} \tilde{c}_6^{B_d}
                   + {\cal O} \left( \frac{1}{m_b^3},  \frac{16 \pi^2}{m_b^4}\right) \right]
 ,
\nonumber \\
\label{HQE4}
 \end{eqnarray}
with
\begin{eqnarray}
\tilde{c}_6^{B_d} & = & |V_{ud}|^2 \left( F^u B_1 + F^u_S B_2 + G^u \epsilon_1 + G^u_S \epsilon_2 \right)
\nonumber
\\
&+& |V_{cd}|^2 \left( F^c B_1 + F^c_S B_2 + G^c \epsilon_1 + G^c_S \epsilon_2 \right) \; .
\end{eqnarray}
The size of the third contribution in Eq.(\ref{HQE4})
is governed by size of $\tilde{c}_6$ and its pre-factor.
      The pre-factor gives
      \begin{equation}
      \frac{16 \pi^2 f_{B_d}^2 M_{B_d}}{m_b^3} \approx 0.395 \approx  0.05 \, c_3 \; ,
      \end{equation}
      where we used $f_{B_d} = (190.5 \pm 4.2)$ MeV \cite{Colangelo:2010et} for 
      the decay constant. If  $\tilde{c}_6$
      is of order $1$, we would expect corrections of the order of $5 \%$ to the total decay rate, which
      are larger than the formally leading $1/m_b^2$-corrections.
      The LO-QCD expression for $\tilde{c}_6^{B_d}$ can be written as
      \begin{eqnarray}
      \tilde{c}_6^{B_d} & = & |V_{ud}|^2 (1-z)^2 \left\{ \left(3 C_1^2 + 2 C_1 C_2 + \frac13 C_2^2 \right)
                        \left[ (B_2 - B_1) + \frac{z}{2} (4 B_2 - B_1) \right]
                 \right . 
             \nonumber \\
       && \left. \hspace{4cm}+ 2 C_2^2 \left[ (\epsilon_2 - \epsilon_1) + 
                            \frac{z}{2} (4 \epsilon_2 - \epsilon_1) \right]
                  \right\} \; .
      \label{tildec6}
     \end{eqnarray}
     However, in Eq.(\ref{tildec6}) several cancellations are arising. 
     In the first line there is a strong cancellation 
     among the bag parameters $B_1$ and $B_2$. In vacuum insertion approximation $B_1 - B_2$ is zero and the
     next term proportional to $4 B_2 - B_1$ is suppressed by $z \approx 0.055$. 
     Using the latest lattice determination of these parameters  \cite{Becirevic:2001fy}
     - dating back to 2001 - 
     \begin{eqnarray} &&
      B_1 = 1.10 \pm 0.20 \; , \; \; 
      B_2 = 0.79 \pm 0.10 \; , \; \;
      \epsilon_1 = -0.02 \pm 0.02 \; , \; \;
      \epsilon_2 =  0.03 \pm 0.01 \nonumber
      \\
      \label{latticebag}
     \end{eqnarray}
     one finds  $B_1 - B_2 \in [0.01,0.61]$ and  $(4 B_2 - B_1) z/2 \in [0.07, 0.12]$, 
     so the second contribution is slightly suppressed compared to the first one.
     Moreover there is an additional
     cancellation among the $\Delta B =1$ Wilson coefficients. Without QCD the combination 
     $3 C_1^2 + 2 C_1 C_2 + \frac13 C_2^2$ is equal to $1/3$, in LO-QCD this combination 
     is reduced to about 
     $0.05 \pm 0.05$ at the scale of $m_b$ (varying the renormalisation scale between $m_b/2$ and $2 m_b$).
     Hence $B_1$ and $B_2$ give a contribution between 0 and 0.07 to $ \tilde{c}_6^{B_d}$, 
     leading thus at most
     to a correction of about 4 per mille to the total decay rate. This statement depends, however, 
     crucially on the numerical values of the bag parameters, where we are lacking a state-of-the-art 
     determination.
     \\
     There is no corresponding cancellation in the coefficients related to the 
     colour-suppressed bag parameters
     $\epsilon_{1,2}$. According to \cite{Becirevic:2001fy} $\epsilon_2 - \epsilon_1 \in [0.02,0.08]$,
     leading to a correction of at most $1.0 \%$ to the decay rate.
     Relying on the lattice determination in \cite{Becirevic:2001fy} we find that the colour-suppressed
     operators can be numerical more important than the colour allowed operators and the total decay rate
     of the $B_d$-meson can be enhanced by the weak annihilation at most by about $1.4 \%$. The status at
     NLO-QCD will be discussed below.
     \\
     The Pauli interference contribution to the $B^+$-decay rate gives
      \begin{eqnarray}
      \tilde{c}_6^{B^+} & = &  (1-z)^2 
           \left[ \left(C_1^2 + 6 C_1 C_2 + C_2^2 \right) B_1
                   + 6  \left(C_1^2 + C_2^2 \right) \epsilon_1
                  \right] \; .
      \label{tildec6PI}
     \end{eqnarray}
     The contribution of the colour-allowed operator is slightly suppressed by the $\Delta B=1$ Wilson
     coefficients. Without QCD the bag parameter $B_1$ has a pre-factor of one, which changes in LO-QCD
     to about -0.3. 
     Taking again the lattice values for the bag parameter from \cite{Becirevic:2001fy}, we expect 
     Pauli interference contributions proportional to $B_1$ to be of the order of about $-1.8 \%$ of the 
     total decay rate.
     In the coefficient of $\epsilon_1$ no cancellation is arising and we expect 
     (using again \cite{Becirevic:2001fy}) this contribution to be between $0$ and $-1.5\%$ of the total 
     decay rate.
     All in all Pauli interference seems to reduce the total $B^+$-decay rate by about $1.8 \%$ to $3.3\%$.
     The status at NLO-QCD will again be discussed below.
     \\
     In the charm system the pre-factor of the coefficient $c_6$ reads 
      \begin{equation}
      \frac{16 \pi^2 f_{D}^2 M_{D}}{m_c^3} \approx \left\{
      \begin{array}{ccc}
        6.2 \approx 0.6 \, c_3  &\mbox{for} &  D^0, D^+
        \\
        9.2 \approx 0.8 \; c_3 \;     & \mbox{for} &  D_s^+
      \end{array}
      \right.
      \; ,
      \end{equation}
      where we used $f_{D^0}   = ( 209.2 \pm 3.3)$ MeV  and 
                    $f_{D^+_s} = ( 248.3 \pm 2.7)$ MeV 
      \cite{Colangelo:2010et} for the decay constants. Depending on the strength of the
      cancellation among the $\Delta C= 1$ Wilson coefficients and the bag parameters, large 
      corrections seem to be possible now: 
      In the case of the weak annihilation the cancellation of the $\Delta C= 1$ Wilson 
      coefficients seems to be even more pronounced than at the scale $m_b$. Thus a knowledge
      of the colour-suppressed operators is inalienable.
      In the case of Pauli interference no cancellation occurs and we get values for the coefficient
      of $B_1$, that are smaller than $-1$ and we get a sizable, but
      smaller contribution from the colour-suppressed operators.
      Unfortunately there is no lattice determination of the $\Delta C=0$
      matrix elements available, so we cannot make any final, profound statements about the
      status in the charm system. Numerical results for the NLO-QCD case will also be discussed below. 
\\
\\
\underline{\bf Fourth term in Eq.(\ref{HQE}):}
\\
\\
If one takes in the calculation of the weak annihilation and Pauli interference diagrams also
small momenta and masses of  the spectator quark into account, one gets corrections that are
suppressed by four powers of $m_b$ compared to the free-quark decay. These dimension seven 
terms are either given by four-quark operators times the small mass of the spectator quark
or by a four quark operator with an additional derivative.
Examples are the following $\Delta B = 0$ operators
\begin{eqnarray}
  P_1 &=& \frac{m_{d,s}}{m_b} \, \bar{b}_i (1-\gamma_5) d_i \times
        \bar{d}_j (1-\gamma_5) b_j \, ,
\\
P_2 &=& \frac{m_{d,s}}{m_b} \, \bar{b}_i (1+\gamma_5) d_i \times
      \bar{d}_j (1+\gamma_5) b_j  \, ,
\\
P_3 &=& \frac{1}{m_b^2} \, \bar{b}_i \overleftarrow{D}_\rho 
        \gamma_\mu (1-\gamma_5) 
        D^\rho d_i \times
      \bar{d}_j \gamma^\mu (1-\gamma_5) b_j  \, ,
\\
P_4 &=& \frac{1}{m_b^2} \, \bar{b}_i \overleftarrow{D}_\rho  (1-\gamma_5) 
        D^\rho d_i \times 
      \bar{d}_j  (1+\gamma_5) b_j  \, .
\end{eqnarray}
These operators have currently only been estimated within vacuum insertion 
approximation. However, for the corresponding operators appearing in the 
decay rate difference of neutral $B$-meson first studies with QCD sum rules have been
performed \cite{Mannel:2007am,Mannel:2011zza}.
\\
Putting everything together we arrive at the
{\it Heavy-Quark Expansion} of decay rates of heavy hadrons
\begin{eqnarray}
\label{HQE5}
\Gamma  & = & \Gamma_0 + \frac{\Lambda^2}{m_b^2} \Gamma_2 + \frac{\Lambda^3}{m_b^3} \Gamma_3 + \frac{\Lambda^4}{m_b^4} \Gamma_4 + \ldots 
\, ,
\end{eqnarray}
where the expansion parameter is denoted by $\Lambda/m_b$. From the above explanations it is clear
that $\Lambda$ is not simply given by $\Lambda_{QCD}$ - the pole of the strong coupling constant - as 
stated often in the literature. Very naively one expects $\Lambda$ to be of the order of $\Lambda_{QCD}$,
because both denote non-perturbative effects. The actual value of $\Lambda$, has, however, to be determined
by an explicit calculation for each order of the expansion separately. 
At order $1/m_b^2$ one finds that $\Lambda$ is of the order of 
$\mu_\pi$ or $\mu_G$, so roughly below 1 GeV. For the third order $\Lambda^3$ is given by $16 \pi^2 f_B^2 M_B$ times
a numerical suppression factor, leading to values of $\Lambda$ larger than 1 GeV.
Moreover, each of the coefficients $\Gamma_j$, which is a product of a perturbatively calculable Wilson coefficient
and a non-perturbative matrix element, can be expanded in the strong coupling
\begin{eqnarray}
\Gamma_j & = & \Gamma_j^{(0)} + \frac{\alpha_s  (\mu)}{ 4 \pi   } \Gamma_j^{(1)} 
                              + \frac{\alpha_s^2(\mu)}{(4 \pi)^2} \Gamma_j^{(2)} + \dots
 \; .
\end{eqnarray}
Before we apply this framework to experimental observables, we would like to make some comments of caution.
\\
A possible drawback of this approach might be that the expansion in the
inverse heavy quark mass does not converge well enough --- advocated 
under the labelling {\it violation of quark hadron duality}.
There is a considerable amount of literature about theoretical 
attempts to prove or to disprove duality, but all of these attempts 
have to rely on strong model assumptions. 
\\
Kolya published some general investigations of quark hadron duality violation in
\cite{Chibisov:1996wf,Bigi:2001ys} and some investigations within the two dimensional
't Hooft model \cite{Bigi:1998kc,Lebed:2000gm}, that indicated the validity of quark hadron
duality. 
Other investigations in that direction were e.g. performed by Grinstein and Lebed in 
1997 \cite{Grinstein:1997xk} and 1998 \cite{Grinstein:1998gc} and by Grinstein
in 2001 \cite{Grinstein:2001zq,Grinstein:2001nu}.
In our opinion the best way of tackling this question is to confront 
precise HQE-based predictions with precise experimental data.
An especially well suited candidate for this problem is the
decay $b \to c \bar c s$, which is CKM dominant, but phase space suppressed. The actual 
expansion parameter of the HQE is in this case not $1/m_b$ but $1/(m_b \sqrt{1- 4 z})$; so 
violations of duality should be more pronounced.
Thus a perfect observable for testing the HQE is the decay rate
difference $\Delta \Gamma_s$ of the neutral $B_s$ mesons, which
is governed by the $b \to c \bar c s$ transition.
The first measurement of this quantity in 2012 and several follow-up
measurements are in perfect agreement with the HQE prediction and exclude
thus huge violations of quark hadron duality,
see \cite{Lenz:2012mb} and the discussion below.

\subsection{Overview of observables}
In this section we give a brief overview of observables, whose experimental values can
be compared with HQE predictions. As we have discussed above,
the general expression for the lifetime ratio of two heavy hadrons $H_1$ and $H_2$ 
reads
\begin{eqnarray}
\frac{\tau (H_1)}{\tau (H_2)} = \frac{\Gamma_2}{\Gamma_1} = 
                                   1 + \frac{\Gamma_2 - \Gamma_1}{\Gamma_1} 
 = 1 &+ & 
                    \frac{\mu_\pi^2(H_1) - \mu_\pi^2(H_2)}{2m_b^2}
           +   \frac{c_G}{c_3} \frac{\mu_G^2  (H_2) - \mu_G^2  (H_1)}{2m_b^2}
\nonumber
\\
 &+ & 
               \frac{c_6(H_2)}{c_3} \frac{\langle H_2 | Q | H_2 \rangle}{m_b^3 M_B} - 
               \frac{c_6(H_1)}{c_3} \frac{\langle H_1 | Q | H_1 \rangle}{m_b^3 M_B}
+ {\cal O}\left(\frac{ \Lambda^4}{m_b^4}\right)  \; ,
\nonumber
\\
\end{eqnarray}
where we have used  the HQE
expression for $\Gamma_1$ and expanded the ratio consistently in $1/m_b$. 
Another possibility would be to use the experimental value for the lifetime $\tau_1$ of 
the hadron $H_1$ and the relation $\Gamma_1 = 1/\tau_1$ to express
the decay rate $\Gamma_1$. This gives
\begin{eqnarray}
\frac{\tau (H_1)}{\tau (H_2)} 
 = 1 &+ & \frac{G_F^2 m_b^3}{384 \pi^3} V_{cb}^2 \tau_1
          \left[ c_3          \left(\mu_\pi^2(H_1) - \mu_\pi^2(H_2)\right)
             +   c_G          \left(\mu_G^2  (H_2) - \mu_G^2  (H_1)\right) \right]
\nonumber
\\
 &+ & \frac{G_F^2 m_b^2}{192 \pi^3} V_{cb}^2 \tau_1
          \left[ 
               \frac{ c_6(H_2)\langle H_2 | Q | H_2 \rangle - 
               c_6(H_1)\langle H_1 | Q | H_1 \rangle}{M_B} + {\cal O}\left(\frac{ \Lambda}{m_b}\right) 
           \right] \; .
\nonumber
\\
\end{eqnarray}
Both methods yield similar numerical results. The relative difference of them is given by
the deviation of the $b$-lifetime prediction in Eq.({\ref{taubfull}) from the measured lifetime:
\begin{equation}
\delta = \frac{1.65 \, \rm ps}{1.519 \, \rm ps} = 1.086 \; .
\end{equation}
Switching between the two methods will change the relative size of the HQE-corrections
by $9\%$. This intrinsic uncertainty has to be kept in mind for error estimates; it could be reduced by 
an NNLO-QCD calculation of $c_3$.
\\
We will discuss the following classes of lifetime ratios:
\begin{itemize}
\item In the case of B-mesons, there are two well-measured ratios
            \begin{equation}
            \frac{\tau (B_s)}{\tau (B_d)}\, , \hspace{1cm} \frac{\tau (B^+)}{\tau (B_d)}  \; .
            \end{equation}
      We have an almost perfect cancellation in the first ratio, therefore this clean ratio can be used
      to search for new physics effects, see, e.g., \cite{Bobeth:2014rda,Bobeth:2011st}. The second 
      ratio 
      is dominated by Pauli interference.
\item Concerning b-baryons, we expect some visible $1/m_b^2$- and $1/m_b^3$-corrections. 
      Until recently only the $\Lambda_b$ lifetime was studied experimentally. In 2014 
      also more precise numbers for the $\Xi_b$-baryons became 
      available \cite{Aaij:2014esa,Aaij:2014lxa} and we can study now ratios like
            \begin{equation}
            \frac{\tau (\Lambda_b)}{\tau (B_d)} \, ,\hspace{1cm} 
            \frac{\tau (\Xi_b^+)}{\tau (\Xi_b^0)} \; .
            \end{equation}
      \item The $B_c$-meson is quite different from the above discussion, because now both
             constituent quarks have large decay rates and we have simultaneously an expansion
            in $1/m_b$ and in $1/m_c$.
            \begin{equation}
            {\tau (B_c)} \; .
            \end{equation}
      \item The ratio of D-meson lifetimes is similar to the ones of $B$-mesons. 
            The big issue is here simply if
            the HQE shows any convergence at all in the ratios
            \begin{equation}
            \frac{\tau (D_s^+)}{\tau (D^0)} \, ,
             \hspace{1cm} \frac{\tau (D^+)}{\tau (D^0)} \; . 
            \end{equation}
      \end{itemize}
Decay rate differences $\Delta \Gamma$ of neutral mesons can determined by a very similar HQE approach  
as discussed above, see, e.g., \cite{Lenz:2012mb} for an introduction into mixing. The general 
expressions for the mixing contribution $\Gamma_{12}$ starts at order $1/m_b^3$ and it can be written as
\begin{eqnarray}
\Gamma_{12}^q & = & \left( \frac{\Lambda}{m_b} \right)^3 \Gamma_3 + \left(\frac{\Lambda}{m_b} \right)^4 \Gamma_4 + 
 \dots \; .
\end{eqnarray}
In the mixing sector we get the following observables:
      \begin{itemize}
      \item In the neutral $B$-meson system $\Delta \Gamma_q$ denotes the difference of the
            total decay rates of the heavy (H) mesons eigenstate and the light (L) eigenstate.
            They are extracted from $\Gamma_{12}$ via the relations 
            \begin{equation}
            \Delta \Gamma_d = \Gamma_{L}^d - \Gamma_{H}^d = 2 |\Gamma_{12}^d| \cos \phi_d  \; ,
             \hspace{1cm}  
            \Delta \Gamma_s = \Gamma_{L}^s - \Gamma_{H}^s = 2 |\Gamma_{12}^s| \cos \phi_s  \; ,
            \end{equation}
            with the mixing phase defined as $\phi_q = \arg (- M_{12}^q/\Gamma_{12}^q)$.
            Related quantities, that also rely on the HQE for $\Gamma_{12}$  are the 
            so-called semi-leptonic asymmetries
            \begin{equation}
            a_{sl}^d = \left| \frac{\Gamma_{12}^d}{M_{12}^d} \right| \sin \phi_d \; ,\hspace{1cm}  
            a_{sl}^s = \left| \frac{\Gamma_{12}^s}{M_{12}^s} \right| \sin \phi_d \; ,
            \end{equation}
            that were already discussed in 1987 by Bigi, Khoze, Uraltsev and Sanda \cite{Bigi:1987in}
            and even earlier in \cite{Hagelin:1980bt,Hagelin:1981zk,Buras:1984pq}.
           . 
      \item In the case of neutral D-mesons the expression of the decay rate difference 
            $\Delta \Gamma_D$ in terms of $\Gamma_{12}$ and $M_{12}$ is more complicated, than
            in the case of $B$-mesons. Here, typically the quantity $y$ is discussed
            \begin{equation}
            y = \frac{\Delta \Gamma_D}{2 \Gamma_D} \;.
            \end{equation}
      \end{itemize}
Before comparing recent data with HQE predictions, we will do some historical investigations of the 
origin of the HQE. 

\section{A brief history of lifetimes and the HQE}
\label{lifetime_sec3}
We give here a brief history of the theoretical investigations of lifetimes of heavy hadrons and
the heavy quark expansion. We do not discuss the development of the Heavy Quark Effective Theory 
(HQET),
which happened in the late 1980s and early 1990s. We also concentrate on total decay rates, 
thus leaving out many of the important contributions to the theory of semi-leptonic decays.
\\
Heavy hadrons were discovered as $J/ \psi$-states in 1974
\cite{Aubert:1974js,Augustin:1974xw}. At about that time the first investigations of weak 
decays of heavy hadrons started. We structure the theoretical development in three periods: 
pioneering studies,
systematic studies and precision studies. It is of course quite arbitrary, where the exact 
borders between these periods are drawn.
\subsection{Pioneering studies}
      Here we summarise the first investigations of heavy meson decays, without
      having a systematic expansion at hand.
      \begin{itemize}
      \item According to Kolya (see, e.g., \cite{Bigi:2011gf})
            \footnote{In an email from 4.11.2012 Kolya wrote to me: 
{\it The present generation may not appreciate how nontrivial
(or even heretic) such a proposition could sound that time! It was the
era of traditional hadron physics where descriptions like Veneziano
model or Regge theory were assumed to underlie hadrons, and their
common (indisputable) feature was soft interactions leading to
exponential suppression of any form factor...
            } 
            }  the first time, that heavy flavour hadrons 
            have been described asymptotically by a free quark decay was in  
            1973 by Nikolaev \cite{Nikolaev:1973uu}. The charm-quark decay as
            the dominant contribution to $D$-meson decays was considered, e.g., in 1974/5 by
            Gaillard, Lee and Rosner \cite{Gaillard:1974mw}, by
            Kingsley, Treiman, Wilczek and Zee \cite{Kingsley:1975fe}, by
            Ellis, Gaillard and Nanopoulos \cite{Ellis:1975hr} and by
            Altarelli, Cabibbo and Maiani \cite{Altarelli:1975zz}.
            In \cite{Ellis:1975hr} the total lifetime of the charm meson was calculated to be
            about $0.5 \, {\rm ps}^{-1}$, by taking only the LO-QCD value of $c_3$ with vanishing 
            internal quark masses into account.
      \item Pauli interference was introduced in 1979 by Guberina, Nussinov, Peccei and
            R{\"u}ckl \cite{Guberina:1979xw}. Without having any systematic expansion
            at hand these authors found
            \begin{equation}
            \frac{\tau (D^+)}{\tau (D^0)}^{\rm PI \, 1979} = \frac{c_-^2 + 2 c_+^2 + 2}{4 c_+^4 + 2} 
                     = \frac{{\cal N}_a + 2}{{\cal N}_a  + 2 + 
                       \left( C_1^2 + 6 C_1 C_2 + C_2^2 \right)}\; .
            \label{tauDancient}
            \end{equation}
            This result can be obtained from our formulae by the following modifications:
            \begin{itemize}
            \item For the $D^0$ decay rate only $\Gamma_0$, i.e., only the free quark decay, 
                  is taken into account in LO-QCD and
                  with vanishing internal quark masses, i.e., no $1/m_c^2$- and $1/m_c^3$-corrections are 
                  considered.
            \item For the $D^+$ decay rate only $\Gamma_0$ and the Pauli interference in $\Gamma_3$ 
                  are taken into account in LO-QCD and with vanishing internal quark masses. 
                  Since at that time no systematic expansion was available, the contributions were
                  simply added. This corresponds to making the following replacements in our
                  formulae: $(4 \pi f_D)^2 \approx (2.63 \, {\rm GeV})^2 \to m_c^2$ and $M_D \approx m_c$,
                  which is of course very crude and more importantly not really justified. 
                  In addition the bag parameters were used in 
                  vacuum insertion approximation, i.e.,
                  $B= 1$ and $\epsilon = 0$.
            \end{itemize}
            With modern inputs Eq.(\ref{tauDancient}) 
            gives a value of about 1.5, while the authors obtained with
            input parameters from 1979 and without using the renormalisation group for the 
            $\Delta C =1$ Wilson coefficients a 
            ratio of about 10.
            It is also quite interesting to note Fig.~1 of \cite{Guberina:1979xw}, 
            which presents the  
            leading $\bar{c} c$-term, weak annihilation and Pauli interference.
            Further studies of Pauli interference were done slightly later in, e.g., \cite{Kobayashi:1980vg}.
      \item Weak annihilation suffers from chirality suppression, thus it was  
            proposed in 1979 by Bander, Silverman and Soni \cite{Bander:1979jx}
            and also by Fritzsch and Minkowski \cite{Fritzsch:1979sr}
            and by Bernreuther and Nachtmann and Stech \cite{Bernreuther:1980bw} to consider gluon emission 
            from the ingoing quark lines in order to explain the large lifetime ratio in the 
            charm system, see Fig.~\ref{fig:WAgluon}.
            \begin{figure}[htb]
            \centering
            {
            \includegraphics[width=0.28\textwidth]{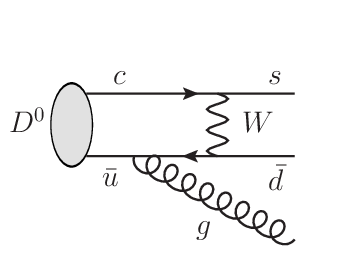}
            \label{fig:WAgluon1}
            }
            \quad
            {
            \includegraphics[width=0.28\textwidth]{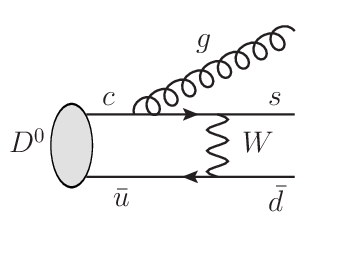}
            \label{fig:WAgluon2}
            }
            \quad
            {
            \includegraphics[width=0.28\textwidth]{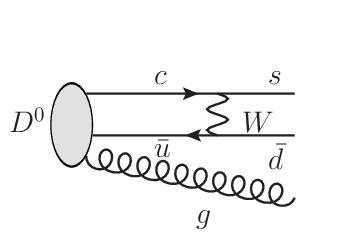}
            \label{fig:WAgluon3}
            }
            \caption{Gluon emission from the weak annihilation diagram.\label{fig:WAgluon}}
            \end{figure}
            This yields a large contribution proportional to $f_D^2/\langle E_{\bar{q}}^2\rangle$, 
            where \(f_D\approx200\text{ MeV}\) is the \(D\) meson decay
            constant and \(\langle E_{\bar{q}} \rangle \) denotes the average energy of the 
            initial anti-quark. Thus the one-gluon emission weak annihilation seems to be not suppressed
            at all, compared to the leading free-quark decay.
            In \cite{Bernreuther:1980bw}, the authors additionally included the 
            Cabibbo-suppressed weak annihilation of \(D^+\) and obtained for the effects
            of weak annihilation in \(D^0\) and \(D^+\)
            \begin{equation}
            \frac{\tau(D^+)}{\tau(D^0)}^{\rm WA \, 1980}  \approx5.6-6.9 \, . 
            \end{equation}
            One should keep in mind, that Pauli interference, 
            which is now known to be the dominant effect, is still neglected here. 
            Comparing with the experimental numbers in the Introduction, one sees what a severe 
            overestimation these early analyses, that did not allow for any power-counting, were.
            If the arguments of \cite{Bander:1979jx,Fritzsch:1979sr,Bernreuther:1980bw} were
            correct, then no systematic HQE would be possible - we come back to this point below.
\item  More systematic studies and further investigations of the Pauli interference effect 
       can be found in
       \cite{Khoze:1983yp,Shifman:1984wx,Bilic:1984nq}. The following formula 
       - Eq.(\ref{HQEorigin1}) - 
       was 
       first derived by Shifman and Voloshin and presented several years later 
       in the review of Khoze and Shifman 
       from 1983 
       \cite{Khoze:1983yp}. It was pointed out in February 1984 by Bilic that in the
        original version there was a sign error, which was corrected 
       in the same year \cite{Shifman:1984wx} by Shifman and Voloshin and shortly afterwards by 
       Bilic, Guberina, Trampetic \cite{Bilic:1984nq}. 
        \begin{equation}
       \begin{aligned}
        \Gamma(D^+)=
        \frac{G_F^2}{2M_D}\langle D^+|&\frac{m_c^5}{64\pi^3}
        \frac{2C_+^2+C_-^2}{3}\bar{c}c+\frac{m_c^2}{2\pi}
        \Big[\left(C_+^2+C_-^2\right)(\bar{c}\Gamma_\mu T^A d)(\bar{d}\Gamma_\mu T^A c)\\
        & +\frac{2C_+^2-C_-^2}{3}(\bar{c}\Gamma_\mu d)(\bar{d}\Gamma_\mu c)\Big]| D^+ \rangle.
       \end{aligned}
       \label{HQEorigin1}
       \end{equation}
       We have rewritten the original expression in the colour-singlet and colour-octet basis 
       commonly used today for \(\Delta C=0\) operators. In order to compare easier with our 
       formulae we can switch from the $C_+$, $C_-$-basis to the $C_1$, $C_2$-basis
       \begin{eqnarray}
       2 C_+^2 + C_-^2 & = & {\cal N}_a \; ,
       \\
         C_+^2 + C_-^2  & = & 2 (C_1^2 + C_2^2) \; ,
       \\
       2 C_+^2 - C_-^2  & = & C_1^2 + 6 C_1 C_2 + C_2^2 \; .
       \end{eqnarray}
       Neglecting weak annihilation, the total decay rate for \(D^0\) is given by the 
       first term in Eq.(\ref{HQEorigin1}). For the bag parameters
       vacuum insertion approximation is used. In early analyses the lifetime ratios were generally 
       underestimated
       \begin{equation}
       \frac{\tau(D^+)}{\tau(D^0)}^{\rm HQE \, 1984}\approx1.5,
       \end{equation}
       which was mainly due to a too small estimate for the decay constant 
       \(f_D\approx160-170\text{ MeV}\). 
       The present value \cite{Colangelo:2010et} of \(f_D=209.2\text{ MeV}\) yields 
       \(\left.\tau(D^+)\middle/\tau(D^0)\right.\approx2.2\), 
       which drastically improves the consistency with experiments. 
       To some extent Eq.(\ref{HQEorigin1})  given  in  \cite{Khoze:1983yp,Shifman:1984wx} 
       can be seen as a starting point for a systematic
       expansion in the inverse of the heavy quark mass.
\item  In 1986 \cite{Shifman:1986mx} Shifman and Voloshin considered for the first time the effects
       of hybrid renormalisation, coming thus much closer to the present state 
       of theory predictions for the ratio of \(D^+\) and \(D^0\) lifetimes.
       Moreover they predicted \cite{Shifman:1986mx}
       \begin{equation}
       \frac{\tau (B_s      )}{\tau (B_d)}^{\rm HQE \, 1986} \approx 1 \; , \; \;
       \frac{\tau (B^+      )}{\tau (B_d)}^{\rm HQE \, 1986} \approx 1.1 \; , \; \;
       \frac{\tau (\Lambda_b)}{\tau (B_d)}^{\rm HQE \, 1986} \approx 0.96 \; , 
       \label{HQE86}
       \end{equation}
       which is amazingly close to current experimental values.
       \\
       In \cite{Shifman:1984wx,Shifman:1986mx}, it was also argued, 
       that \(\tau(D_s^+)\approx\tau(D^0)\), which contradicted the
       experimental situation at that time. In 1986 it was further shown by Guberina,
       R{\"u}ckl and Trampetic \cite{Guberina:1986gd}
       that 
       the HQE was able to 
       correctly reproduce the hierarchy of lifetimes in the charm sector
       \begin{equation}
       {\rm HQE \, 1986:} \; \; \;
       \tau(D^+)>\tau(D^0)>\tau(\Xi_c^+)>\tau(\Lambda_c^+)>\tau(\Xi_c^0)>\tau(\Omega_c^0).
       \end{equation}
%\cite{Shifman:1986sm}
%\bibitem{Shifman:1986sm}
%  M.~A.~Shifman and M.~B.~Voloshin,
%  %``On Annihilation of Mesons Built from Heavy and Light Quark and anti-B0 <---> B0 Oscillations,''
%  Sov.\ J.\ Nucl.\ Phys.\  {\bf 45} (1987) 292
%   [Yad.\ Fiz.\  {\bf 45} (1987) 463].
  %%CITATION = SJNCA,45,292;%%
  %687 citations counted in INSPIRE as of 10 Apr 2014
In 1986 Khoze, Shifman, Uraltsev and Voloshin \cite{Khoze:1986fa} refined the analysis of
\cite{Khoze:1983yp,Shifman:1984wx}, by taking into account weak annihilation and $B$-mixing. 
In particular they found that the decay rate difference in the neutral $B_s$-system may be sizable
\begin{equation}
\frac{\Delta \Gamma_s}{ \Gamma_s}^{\rm HQE \, 1986} \approx 
0.07 \left(\frac{f_{B_s}}{(130 \; {\rm MeV}}\right)^2 \approx 0.22
\, ,
\end{equation}
 where we inserted
the most recent FLAG-average \cite{Colangelo:2010et} for the decay constant.
The authors of \cite{Khoze:1986fa} emphasised also that the weak annihilation effects suggested 
in \cite{Bander:1979jx,Fritzsch:1979sr,Bernreuther:1980bw} formally leads to huge corrections in the
$1/m_q$-expansion, which spoils a systematic expansion. This problem somehow stopped
\footnote{Blok and Shifman stated in 1992 \cite{Blok:1992hw}:
"{\it Probably for this reason the problem of pre-asymptotic
corrections in the inclusive widths has been abandoned for many years.}".} 
further 
work in that direction until the issue was settled in
January 1992 by Bigi and Uraltsev \cite{Bigi:1991ir}.
      \end{itemize}
\subsection{ Systematic studies}
Here we describe the development of the HQE in its current form.
      \begin{itemize} 
      \item For inclusive semi-leptonic decays, where the above issue was not severe, 
            it was shown already in 1990 by Chay, Georgi and Grinstein \cite{Chay:1990da}, 
            that in an expansion in inverse powers of the heavy quark mass no 
            $1/m_q$-corrections are appearing and therefore a consistent, systematic expansion
            seemed to be in reach for these decays\footnote{The famous Luke's Theorem 
            \cite{Luke:1990eg} 
            was proven in the context of the HQET. This theorem can be considered as a 
            generalisation of the Ademollo-Gatto theorem from 1964 \cite{Ademollo:1964sr}.}.
      \item In 1992 Bigi and Uraltsev \cite{Bigi:1991ir} explained the apparent contradiction between 
            the \(1/m_q\) scaling of the HQE
            and the \(f_D^2/\langle E_{\bar{q}} \rangle^2\) enhanced gluon bremsstrahlung of 
            \cite{Bander:1979jx,Fritzsch:1979sr,Bernreuther:1980bw}. They showed, that
            these power-enhanced terms cancel in fully inclusive rates between different 
            cuts as indicated in Figure \ref{fig:WAcuts} and pre-asymptotic
            effects hence scale with \(1/m_c^3\), consistently with the HQE.
            \begin{figure}[htb]
            \centering
            \includegraphics[width=0.4\textwidth]{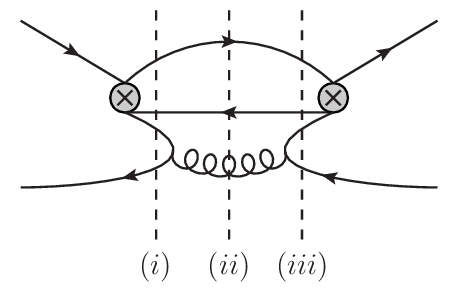}
            \caption{Different cuts contributing to the weak annihilation. The 
            \(f_D^2/\langle E_{\bar{u}} \rangle^2\) enhanced term due to the cut (ii) considered in
            \cite{Bander:1979jx,Fritzsch:1979sr,Bernreuther:1980bw}
            is cancelled by interference effects (i) and (iii), such that the fully inclusive 
            rate experiences the
            correct \(1/m_c^3\) scaling behaviour predicted by the HQE.}
            \label{fig:WAcuts}
            \end{figure}
            This seminal work opened now the way for the HQE in its current form. The 
            explicit proof of the cancellation  of all power-enhanced terms was done in 1998
            by Beneke, Buchalla, Greub, Lenz and Nierste \cite{Beneke:1998sy},
            in the context of the calculation of $\Gamma_3^{(1)}$ for $\Delta \Gamma_s$. 
      \item The HQE in its current form was written down in July 1992 in \cite{Bigi:1992su} by 
            Bigi, Uraltsev and Vainshtein for semi-leptonic and non-leptonic decays with
            one heavy quark in the final state. By working out the expansion in Eq.(\ref{bbarbME}) 
            the absence of $1/m_q$-corrections  was shown also for the non-leptonic decays. 
            In addition $\Gamma_2^{(0)}$ was determined with the inclusion of charm mass effects,
            i.e., the values of the Wilson coefficient $c_5$ given in Eq.(\ref{c5sl}) 
            and Eq.(\ref{c5nl}). In the original
            paper there are some misprints, that were partly corrected\footnote{In Eq.(4) of the
            erratum the factors $m_Q^2$ should be in the denominator instead of the numerator.} 
            in an erratum.
            The full set of correct formulae was given end of 1992 by Bigi, Blok, Shifman, Uraltsev and 
            Vainshtein in \cite{Bigi:1992ne}. In these two papers \cite{Bigi:1992su,Bigi:1992ne}
            a different normalisation was used
            for the physical meson states than we did in Eq.(\ref{bbarbME}).
            \\
            At about the same time Blok and Shifman investigated the rule
            of discarding terms of order $1/N_c$ in inclusive $b \to c \bar{u} d$- and
            $c \to s \bar{d} u$-decays \cite{Blok:1992hw}, as well as in the
            $b \to c \bar{c} s$-decay \cite{Blok:1992he}. In that respect they also determined
            the $1/m_b^2$-corrections for inclusive non-leptonic decays.
            More precisely they determined the contribution of $c_{5,b}^{c\bar{u}d}$ proportional
            to $C_1 C_2$ - see  Eq.(\ref{c5nl}) - in  \cite{Blok:1992hw} and the 
            contribution of $c_{5,b}^{c\bar{u}d}$ 
            proportional to $C_1 C_2$ - see  Eq.(\ref{c5nl2})- in  \cite{Blok:1992he}.
            \\
            The complete formulae for the case of two heavy particles in the final state
            with identical masses, e.g., $b \to c \bar c s$ -  see  Eq.(\ref{c5nl2}) -
            are given in December 1993 by  Bigi, Blok, Shifman and Vainshtein
            and in January 1994 in a book contribution of Bigi, Blok, Shifman, Uraltsev and 
            Vainshtein from 1994 \cite{Bigi:1994wa}. 
            In these papers now the same normalisation
            for the meson states as in  Eq.(\ref{bbarbME}) is used.
            The case for two arbitrary masses was studied by Falk, Ligeti, Neubert and Nir
            \cite{Falk:1994gw} in 1994.
    \item   Now the door was open for many phenomenological investigations, which led also
            to several challenges for the new theory tool: 
            \begin{itemize}
            \item Inclusive non-leptonic
                 decays were considered by Palmer and Stech in May 1993 \cite{Palmer:1993kx}. It turned
                 out that 
                 the theory prediction for the decay $b \to c \bar{c} s$ did not fit to the data.
                 Related investigations of the {\it missing charm puzzle} and the inclusive semi-leptonic
                 branching ratio were done in  November 1993 by Bigi, Blok, Shifman and 
                 Vainshtein \cite{Bigi:1993fm}
                 ({\it The baffling semi-leptonic branching ratio}). 
                 In May 1994 it was suggested by Dunietz,
                 Falk and Wise \cite{Falk:1994hm} ({\it Inconclusive inclusive nonleptonic $B$ decays})
                 that this discrepancy points towards a violation of 
                 local quark hadron duality in the decay $b \to  c \bar{c} s$ - a suggestion, 
                 which is now ruled out by the 2012 measurement of $\Delta \Gamma_s$, which is in perfect 
                 agreement with the HQE prediction, see below.
                 Moreover the current theory prediction for the semi-leptonic branching ratio
                 in Eq.(\ref{Bsltheo}) agrees well with the experimental numbers given in  
                 Eq.(\ref{Bslexp}), although there is still some space for deviations.
            \item An early extraction of $V_{cb}$, $m_c$ and $m_b$ was done in 1993 by Luke and Savage
                  \cite{Luke:1993za} and by Bigi and Uraltsev \cite{Bigi:1993ey} and 
                 further in 1995 by Falk, 
                  Luke and Savage \cite{Falk:1995kn}. This kind of studies form a big industry
                  now, see, e.g., the review about the determination of $V_{cb}$ and $V_{ub}$
                  by Kowalewski and Mannel in the PDG \cite{PDG}.
            \item Bigi and Uraltsev applied the HQE to charm lifetimes in \cite{Bigi:1993ey}
                  and also some aspects in \cite{Bigi:1993bh}.
                  For the \(D_s^+\) meson they found
                  \begin{equation}
                  \frac{\tau(D_s^+)}{\tau(D^0)}^{\rm HQE 1994}=0.9-1.3, 
                  \end{equation}
                  where the uncertainty dominantly arises from the weak annihilation. 
            \item Lifetimes of $b$-hadrons were also further studied.
                  In that respect the expressions for $\Gamma_3^{(0)}$ with charm quark mass 
                  dependence were presented by Kolya Uraltsev \cite{Uraltsev:1996ta} in 1996
                  \footnote{In an email dated from 4.11.2012, Kolya claimed that these results 
                      were known since a long time:            
                {\it Effects of the internal quark masses were in fact considered; the
                expressions were at hand, and plugging numbers were so a simple matter
                that this was not even noted specially. The expressions (they are given,
                for instance, in arXiv:hep-ph/9602324)
                were taken from the same mid-1980s notes I mentioned above.}} and slightly later 
                by Neubert and Sachrajda \cite{Neubert:1996we}. This mass dependence turned out to 
                be important. Moreover the inclusion of colour-suppressed four quark operators
                was found to be crucial.
                Neubert and Sachrajda \cite{Neubert:1996we} gave a very nice and 
                comprehensive review of the status quo 
                in 1996 for the different $b$-hadron lifetime predictions in LO-QCD.
                At that time the measured $\Lambda_b$-lifetime was in conflict with early HQE
                predictions, that predicted a value of around $1.5$  ps, see Eq.(\ref{HQE86}).
                The old data \cite{Buskulic:1992nu,Buskulic:1995xt,Barate:1997if,Ackerstaff:1997qi}
                pointed, however,  more to values around $1.0 - 1.3$ ps.
                \begin{equation}
                \begin{array}{|l|l|l|c|l|}
                \hline
\mbox{Year} & \mbox{Exp} & \mbox{Decay} & \tau (\Lambda_b) \left[ \mbox{ps}\right]& \tau (\Lambda_b)/\tau (B_d)
\\
\hline
\hline
1998        & \mbox{OPAL} & \Lambda_c l      & 1.29 \pm 0.25 & 0.85 \pm 0.16
\\
\hline
1997        & \mbox{ALEPH} & \Lambda_c l     & 1.21 \pm 0.11 & 0.80 \pm 0.07 
\\
\hline
1995        & \mbox{ALEPH} & \Lambda_c l     & 1.02 \pm 0.24 & 0.67 \pm 0.16 
\\
\hline
1992        & \mbox{ALEPH} & \Lambda_c l     & 1.12 \pm 0.37 & 0.74 \pm 0.24 
\\
\hline
\end{array}
\label{tauLambdabold}
\end{equation}
Neubert and Sachrajda concluded that this points - if the experimental values stay - either to
anomalously large matrix elements (they will be discussed below) or to a violation of
quark-hadron duality. The latter attitude was quite popular at that time, 
see, e.g., the paper by Altarelli, Martinelli, Petrarca and Rapuano from 1996 \cite{Altarelli:1996gt}
or the paper from Cheng from 1997 \cite{Cheng:1997xba}
or the work from Ito, Matsuda and Matsui also from 1997 \cite{Ito:1997qq}.
Nowadays we know that the $\Lambda_b$-lifetime was a purely experimental problem and the 
measured values are in good agreement with the HQE estimates. These estimates suffer, however,
from sizable hadronic uncertainties, which could be reduced by a state of the art lattice calculation.
            \item The lifetime of the $B_c$-meson, where both the $b$- and the $c$-quark decay weakly
                  was studied systematically in 1996 by Beneke and Buchalla \cite{Beneke:1996xe}.
            \end{itemize}
      \end{itemize}
\subsection{ Precision studies}
Some motivation and some topics of precision studies can be found already in
the recommendations given in the seminal 1992 paper \cite{Bigi:1992su}:
''{\it The general procedure outlined above can be improved in four respects:
\begin{itemize}
\item[(i)] Some of the numerical predictions stated above were tentative since not all the relevant 
           calculations have been performed yet. Since the ``missing'' computations involve perturbation
           theory this presents ``merely'' a technical delay.
\item[(ii)] The real accuracy obtainable in this approach can be determined by calculating terms of order 
           $1/m_Q^4$ and estimating the size of the relevant matrix elements.
\item[(iii)] ...''
\end{itemize}}
Item (i) concerns the field of determining higher order QCD-corrections. After being involved in several 
NLO-QCD calculations within the HQE, I of course disagree with the use of the word ``{\it merely}'' above. 
Besides being a tedious task, these efforts had also a conceptual value, since they provided an explicit
proof of the arguments for a cancellation of singularities due to quark thresholds given by 
Bigi and Uraltsev \cite{Bigi:1991ir}.
Item (ii) suggests the discussion of higher order terms in HQE, which has been done currently for many 
observables, see below.
Another crucial topic, that was, however, not emphasised in  \cite{Bigi:1992su}, is the non-perturbative
determination of the arising matrix elements. 
\subsubsection{ NLO-QCD} 
            For semi-leptonic decays the NLO QCD corrections in $\Gamma_2^{(1)}$ 
            proportional to $\mu_\pi^2$ 
            were determined in 2007 by Becher, Boos and Lunghi \cite{Becher:2007tk} and 
            confirmed in 2012 \cite{Alberti:2012dn}.
            The corresponding corrections proportional to $\mu_G^2$ were calculated 
            very recently by Alberti, 
            Gambino and Nandi \cite{Alberti:2013kxa}.
            \\           
            As discussed above, the NLO QCD corrections in $\Gamma_3^{(1)}$ were crucial for 
            proofing the consistency of the HQE. They were determined 
            for $\Delta \Gamma_s$ in 1998 by Beneke, Buchalla, Greub, Lenz and 
            Nierste \cite{Beneke:1998sy}. In this case the diagrams in Fig.(\ref{fig:NLO}) 
            appear.
            \begin{figure}[htb]
            \centering
            \includegraphics[width=0.5\textwidth, angle =270]{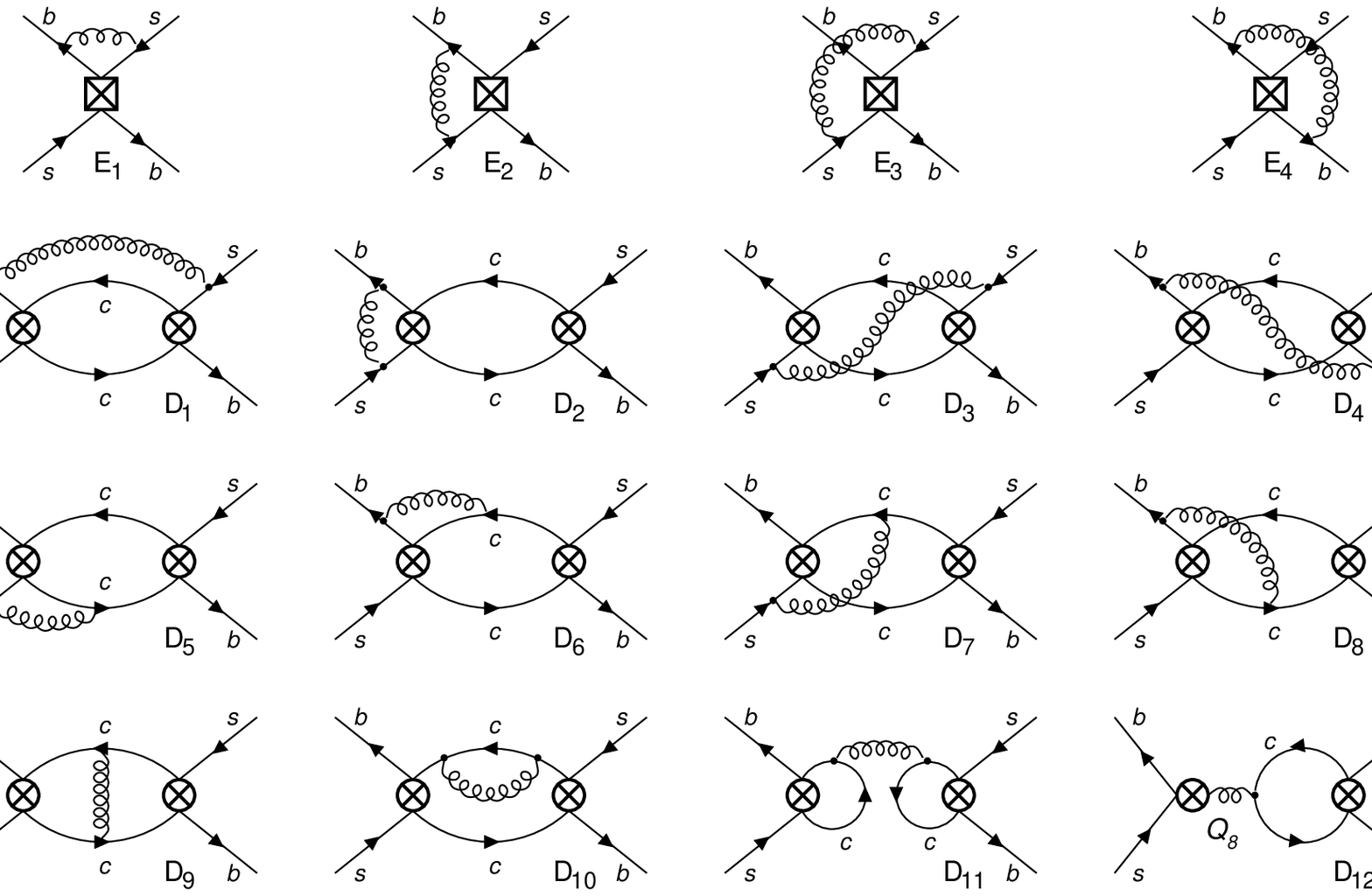}
            \caption{NLO-QCD diagrams contributing to $\Delta \Gamma_s$.}
            \label{fig:NLO}
            \end{figure}
            The same authors as well as the Rome group - Franco, Lubicz, Mescia and Tarantino -
            calculated the NLO-QCD corrections for $\tau (B+)/\tau (B_d)$ in 
            2002 \cite{Beneke:2002rj,Franco:2002fc}. The Rome group included also the NLO-corrections
            for $\tau (B_s)/\tau (B_d)$ and partly for $\tau (\Lambda_d)/\tau (B_d)$ 
            \cite{Franco:2002fc} - here some penguin diagrams are still missing.
            Some dominant NLO-corrections for  $\tau (B_s)/\tau (B_d)$ have already been determined in 
            1998 by Keum and Nierste \cite{Keum:1998fd}.
            In \cite{Beneke:2002rj} it was also shown that the use of $\bar{m}_c (\bar{m}_b)$ 
            automatically sums up logarithms of the form $\alpha_s^n z \log^n z$ to all orders.
            For $\Delta \Gamma_d$ and the semi-leptonic asymmetries $a_{sl}^d$ and $a_{sl}^s$,
            $\Gamma_3^{(1)}$ was determined in 2003 by Beneke, Buchalla, Lenz and Nierste
            \cite{Beneke:2003az} and by Ciuchini, Franco, Lubicz, Mescia and Tarantino 
            \cite{Ciuchini:2003ww}.
            For the decay rate difference of neutral $D$-mesons the above formulae were rewritten
            in 2010 by Bobrowski, Lenz, Riedl and Rohrwild \cite{Bobrowski:2010xg}
            and for D-meson lifetime ratios some missing contributions were calculated in 2013
            by Lenz and Rauh \cite{Lenz:2013aua}.
            \\
            A pioneering studies of some integrals that appear in NNLO-QCD for $\Delta \Gamma_s$ 
            has been performed in 2012 by Asatrian, Hovhannisyan and Yeghiazaryan 
            \cite{Asatrian:2012tp}.

\subsubsection{Higher order terms in the OPE}
            For semi-leptonic decays $1/m_b$ corrections to the kinetic and chromo-magnetic 
            operator were studied in 1994 by Bigi, Shifman,  Uraltsev and 
            Vainshtein \cite{Bigi:1994ga}.
            Similar contributions to semi-leptonic decays were studied in  
            1995 by Blok, Dikeman and Shifman \cite{Blok:1994cd}
            and
            1996 by Kremm and Kapustin \cite{Gremm:1996df}. 
            Even higher corrections - $\Gamma_4^{(0)}$ and $\Gamma_5^{(0)}$ - 
            to the semi-leptonic decay width were investigated in  2010 by 
            Mannel, Turczyk and Uraltsev
            \cite{Mannel:2010wj}.
            \\
            $1/m_q$-corrections to weak annihilation and Pauli interference, i.e.,
            $\Gamma_4^{(0)}$ were determined for $\Delta \Gamma_s$ in 1996 by Beneke, Buchalla and
            Dunietz \cite{Beneke:1996gn} and they turned out to be sizable.
            The corresponding corrections for  $\Delta \Gamma_d$ were calculated 
            in 2001 by Dighe, Hurth, Kim and Yoshikawa \cite{Dighe:2001gc} 
            and for $b$-lifetimes in 
            by Gabbiani, Onishchenko and Petrov in 2003 \cite{Gabbiani:2003pq} 
            and 2004 \cite{Gabbiani:2004tp} (in  the latter one also $1/m_q^2$-corrections 
            were investigated) and by Lenz and Nierste in 2003 \cite{LN2003}. 
            Badin, Gabbiani and Petrov studied also $\Gamma_5^{(0)}$ for $\Delta \Gamma_s$ in 
            2007 \cite{Badin:2007bv}. In  $\Gamma_5^{(0)}$ several  completely unknown matrix 
            elements are arising. Moreover the Wilson coefficients have very small numerical values.
            Thus we are not including these corrections in our estimates.
            \\
            One can also try to apply the methods of the HQE to $D$-mixing. First efforts
            in that direction were made in 1992 by Georgi \cite{Georgi:1992as} and by
            Ohl, Ricciardi and Simmons \cite{Ohl:1992sr}. It turns out that the leading term, 
            $\Gamma_3$, suffers from a severe GIM cancellation and thus the HQE leads to very small
            predictions for $D$-mixing. One idea to circumvent this severe cancellation was 
            to consider
            higher orders in the HQE, in particular $\Gamma_6$ and $\Gamma_9$.
            Bigi and Uraltsev have shown in 2000 \cite{Bigi:2000wn} how in  $\Gamma_6$ and $\Gamma_9$
            the $1/m_c$-suppression could be overcompensated by a lifting of the GIM-suppression.
            They concluded that values of $x$ and $y$ of up to $1 \%$ are not excluded within the 
            HQE.
\subsubsection{Non-perturbative parameters}
            Early studies of  $\mu_\pi^2$ have been done, e.g., in 1993 by Bigi, Shifman, Uraltsev 
            and Vainshtein
            \cite{Bigi:1993ex}.  In 1994 \cite{Bigi:1994re} some ideas how to extract this quantity
            from experiment were developed by  Bigi, Grozin, Shifman, Uraltsev and Vainshtein.
            The same quantity has also been determined with QCD sum rules in 1993
            by Ball and Braun \cite{Ball:1993xv}. A kind of contradicting result was obtained in
            1996 by Neubert \cite{Neubert:1996wm} with the same method. Calculations within lattice
            QCD were, e.g., performed by Kronfeld and Simone in 2000 \cite{Kronfeld:2000gk}. 
            The most recent value for
            $\mu_\pi^2$ for $B$-mesons comes from a fit of semi-leptonic decays by Gambino and 
            Schwanda in 2013 \cite{Gambino:2013rza}, somehow confirming the first QCD sum rule 
            calculation: 
            \begin{equation}
            \begin{array}{|c|l|}
            \hline
            \mu_\pi^2 & 
            \\
            \hline
            \hline
            (0.52 \pm 0.12) \; {\rm GeV}^2 & {\rm QCD-SR \, 1993}
            \\
            \hline
            (0.10 \pm 0.05) \; {\rm GeV}^2 & {\rm QCD-SR \, 1996} 
            \\
            \hline
            (0.45 \pm 0.12) \; {\rm GeV}^2 & {\rm Lattice \, 2000}
            \\
            \hline
            (0.414 \pm 0.078) \; {\rm GeV}^2 & {\rm Fit \, 2013} 
            \\
            \hline
            \end{array}
            \end{equation}
            $\mu_G^2$ can in principle be determined from experiment - see Eq.(\ref{muG}) -,
            for B-mesons it was further investigated by Kolya in 2001 \cite{Uraltsev:2001ih}.
            The differences of $\mu_G^2$ and $\mu_\pi^2$, if one considers instead of the lightest 
            $B$-mesons, $B_s$-mesons or $\Lambda_b$-baryons were studied by 
            Bigi, Mannel and Uraltsev in 2011 \cite{Bigi:2011gf}.
            \\
            The new results seem also to confirm the bound 
            \begin{equation}
            \mu_\pi^2 > \mu_G^2
            \end{equation}
            that was derived by Bigi, Shifman, Uraltsev 
            and Vainshtein \cite{Bigi:1993ex,Bigi:1994ga}
            and by Voloshin \cite{Voloshin:1994pa}.  Kapustin, Ligeti, Wise and 
            Grinstein \cite{Kapustin:1996dy} claimed that the above bound will be weakened due to
            perturbative corrections. A study of Kolya Uraltsev \cite{Uraltsev:1996rd} came,
            however, to a different conclusion.
            \\
            Matrix elements of four-quark operators relevant for lifetime ratios are almost unknown.
            For $\tau (B^+) / \tau (B_d)$ the latest lattice calculation was performed by Becirevic 
            in 2001 and only published as proceedings \cite{Becirevic:2001fy}. Unfortunately these
            parameters, see Eq.(\ref{latticebag}) have never been updated. The same matrix
            elements can also be used for the case of  $\tau (B_s) / \tau (B_d)$.
            There is also an earlier lattice study from Di Pierro and Sachrajda from 1999 
            \cite{DiPierro:1998ty}, as well as two QCD sum rule studies from Baek, Lee, Liu and Song
            in 1997 \cite{Baek:1998vk} and one year later from Cheng and Yang \cite{Cheng:1998ia}.
            \begin{equation}
            \begin{array}{|c|c|c|c||l|}
            \hline
            B_1          &     B_2     &\epsilon_1    &\epsilon_2    &
            \\ 
            \hline
            \hline
            1.01 \pm 0.01&0.99 \pm 0.01&-0.08 \pm 0.02&-0.01 \pm 0.03&1997 \; {\rm QCD-SR}
            \\
            \hline
            1.06 \pm 0.08&1.01 \pm 0.06&-0.01 \pm 0.03&-0.01 \pm 0.02&1998 \; {\rm Lattice}
            \\
            \hline
            0.96 \pm 0.04&0.95 \pm 0.02&-0.14 \pm 0.01&-0.08 \pm 0.01&1998 \; {\rm QCD-SR}
            \\
            \hline
            1.10 \pm 0.20&0.79 \pm 0.10&-0.02 \pm 0.02& 0.03 \pm 0.01&2001 \; {\rm Lattice}
            \\
            \hline
            \end{array}
            \label{BagBmeson}
            \end{equation}
            Comparing these numbers, the authors \cite{Baek:1998vk,Cheng:1998ia}
            of the QCD sum rule evaluation seem to have very aggressive error estimates. 
            Because of the very pronounced cancellations in Eq.(\ref{tildec6}) precise values 
            for these bag parameters are crucial for an investigation of the lifetime ratio
            $\tau (B^+) / \tau (B_d)$.
            To some extent our definition of the bag parameters given in Eq.(\ref{Bag1}) and
            Eq.(\ref{Bag2}) above was a little too simplistic.
            In reality we are considering the isospin breaking combinations
            \begin{eqnarray}
            \frac{\langle B_d |Q^d -Q^u  | B_d \rangle}{M_{B_d}} =  f_B^2 B_1 M_{B_d} \; ,
            &&
            \frac{\langle B_d |Q^d_S  -Q^u_S| B_d \rangle}{M_{B_d}} = f_B^2 B_2 M_{B_d} \; ,
            \\
            \label{Bag3}
            \frac{\langle B_d |T^d   -T^u | B_d \rangle}{M_{B_d}} =  f_B^2 \epsilon_1 M_{B_d} \; ,
            &&
            \frac{\langle B_d |T^d_S  -T^u_S| B_d \rangle}{M_{B_d}} =  f_B^2 \epsilon_2 M_{B_d} \; .
            \label{Bag4}
            \end{eqnarray}
            This definition leads to the cancellation of unwanted penguin contractions, 
            see Fig.~\ref{fig:peng} and enables
            thus in principle very precise calculations.
            \begin{figure}[htb]
            \centering
            \includegraphics[width=0.7\textwidth]{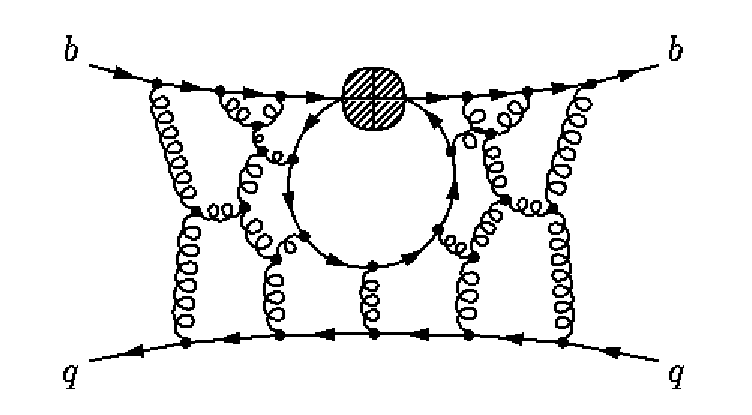}
            \caption{Penguin contractions that cancel in lifetime ratios like $\tau (B^+) / \tau (B_d)$
                     and $\tau (\Xi_b^0) / \tau (\Xi_b^+)$. They will, however, give a contribution
                     to  $\tau (\Lambda) / \tau (B_d)$. Since these contributions introduce a mixing
                     of operators of different dimensionality, they are difficult to handle.}
            \label{fig:peng}
            \end{figure}
            For the $\Lambda_b$-lifetime our knowledge of the matrix elements is even worse. 
            There is only an exploratory
            lattice study from Di Pierro, Sachrajda and Michael available, dating back to 1999
            \cite{DiPierro:1999tb}. Here also any update would be extremely welcome.
            In that case two matrix elements are arising that are parameterised by $L_1$ and $L_2$
            \begin{eqnarray}
            %\frac{\langle \Lambda_b | \bar{b} \gamma^{\mu} (1-\gamma_5) q \cdot 
            %        \bar{q} \gamma_{\mu} (1-\gamma_5) b | \Lambda_b\rangle}
            %        { M_{\Lambda_b}} 
            %= 
            \frac{\langle \Lambda_b | Q^q | \Lambda_b\rangle}{M_{\Lambda_b}} 
            & = & 
            f_B^2 M_B L_1\, ,
            \\
            %\frac{\langle \Lambda_b | \bar{b} \gamma^{\mu} (1-\gamma_5) T^a q \cdot 
            %        \bar{q} \gamma_{\mu} (1-\gamma_5) T^a b  | \Lambda_b\rangle}
            %      {M_{\Lambda_b}}
            %= 
            \frac{\langle \Lambda_b | T^q | \Lambda_b\rangle}{M_{\Lambda_b}} 
            & = & 
            f_B^2 M_B L_2  \, ,
            \end{eqnarray}
            where the operators were defined in Eq.(\ref{operator1}) and Eq.(\ref{operator2}).
            The numerical values obtained in \cite{DiPierro:1999tb} are shown in Eq.(\ref{rnum}).
            In elder works the colour re-arranged operator was investigated instead of the
            colour octett operator $T^q$. There the following definition was used
            \begin{eqnarray}
            \frac{\langle \Lambda_b | \bar{b}_\alpha \gamma^{\mu} (1-\gamma_5) q_\alpha \cdot 
                    \bar{q}_\beta \gamma_{\mu} (1-\gamma_5) b_\beta \rangle}
                    {M_{\Lambda_b}} 
            & = & 
            - \frac{f_B^2 M_B}{6} r\, ,
            \\
            \frac{\langle \Lambda_b | \bar{b}_\alpha \gamma^{\mu} (1-\gamma_5) q_\beta \cdot 
                    \bar{q}_\beta \gamma_{\mu} (1-\gamma_5) b_\alpha \rangle}
                  {M_{\Lambda_b}}
            & = & 
            \tilde{B} \frac{f_B^2 M_B}{6} r  \, .
            \end{eqnarray}
            The two parameter sets are related by\footnote{In \cite{DiPierro:1999tb} the parameters
            $r$ and $\tilde{B}$ were interchanged in Eq.(\ref{Umr}), while the correct relation was
            given in \cite{DiPierro:1998cj}.}
            \begin{eqnarray}
             r = - 6 L_1 \, , && \tilde{B} = - \frac13 - 2 \frac{L_2}{L_1} \, .
            \label{Umr}
            \end{eqnarray}
            Because of the long standing discrepancy between experiment and theory for the 
            $\Lambda_b$-lifetime people of course tried different methods to determine the missing
            matrix elements:
            Rosner related in 1996 the four-quark matrix elements to results from 
            spectroscopy \cite{Rosner:1996fy} and found\footnote{Neubert and 
            Sachrajda \cite{Neubert:1996we} quoted
            in 1996 this formula as 
            $ r = 4/3 (M^2_{\Sigma_b^*} - M^2_{\Sigma_b})/(M^2_{B^*} - M^2_B)$,
            which gives values that are about $10 \%$ larger than Eq.(\ref{Rosner}). 
            Cheng \cite{Cheng:1997xba} quoted in 1997 the same formula as given in Eq.(\ref{Rosner}).}
            \begin{equation}
             r = \frac43 \frac{M_{\Sigma_b^*} - M_{\Sigma_b}}{M_{B^*} - M_B} \; .
            \label{Rosner}\end{equation}
            At that time the values of the masses of the baryons were almost unknown, 
            which resulted in
            quite rough and large estimates, yielding $r \approx 1.6$.
            This situation changed completely now
            and we will use the method of Rosner with new experimental numbers for the 
            baryon masses\footnote{PDG \cite{PDG} gives: 
            $ M_{\Sigma_b^*} - M_{\Sigma_b}  =   21.2  (2.0)$ MeV,
            $ M_{\Sigma_b^*}                 = 5832.1  (2.0)$ MeV, 
            $ M_{\Sigma_b  }                 = 5811.3  (2.0)$ MeV
            and
            $ M_{B^*}                        =  5325.20 (0.40)$ MeV.}
            to update the $\Lambda_b$ lifetime below.
            Colangelo and de Fazio applied in 1996 \cite{Colangelo:1996ta}
            the method of QCD sum rules and obtained relatively small numbers for $r$. 
            Huang, Liu and Zhu managed in 1999 \cite{Huang:1999xj}, however, to obtain with the 
            same method much larger numbers, that also lead to a lifetime of the $\Lambda_b$-baryon,
            that was compatible with the measurements at that time.
            Even earlier (1979) estimates within the bag model and the non-relativistic quark model 
            for charmed hadrons from 
            Guberina, Nussinov, Peccei and R{\"u}ckl \cite{Guberina:1979xw} pointed towards smaller values
            of $r$. All in all currently the following numerical values are available:
            \begin{equation}
            \begin{array}{|c|c|c|c||l|} 
             \hline             
             L_1 = - \frac{r}{6}     & L_2 \, \mbox{or} \;   \frac{r}{9}    & r            & \tilde{B} &
             \\
             \hline
             \hline
            -0.103(10)    & 0.069(7)    & 0.62(6)     & 1  & 2014 \; {\rm Spectroscopy \, update}
             \\
             \hline
             -0.22(4)     & 0.17(2)     & 1.32(24)    & 1.21(34)  & 1999 \; {\rm Exploratory \, Lattice}
             \\
            \hline
            -0.22(5)      & 0.14(3)     & 1.3(3)      & 1  & 1999 \; {\rm QCD-SR \; v1}
             \\
            -0.60(15)     & 0.40(10)    & 3.6(9)      & 1  & 1999 \; {\rm QCD-SR \; v2}
             \\
             \hline
            -0.033(17)    & 0.022(11)   & 0.2(1)      & 1  & 1996 \; {\rm QCD-SR}
             \\
             \hline
            \approx -0.03 & \approx 0.02& \approx 0.2 & 1  & 1979 \; {\rm Bag \, model}
             \\
             \hline
            \approx -0.08 & \approx 0.06& \approx 0.5 & 1  & 1979 \; {\rm NRQM }
             \\
             \hline
            \end{array}
            \label{rnum}
            \end{equation}
            In \cite{DiPierro:1999tb} the two parameters $L_1$ and $L_2$ were calculated, else only 
            $r$ was determined. In the latter case we assumed $\tilde{B} = 1$ 
            (valence quark approximation)
            in order to determine $L_2$. 
            Comparing all these numbers we find that two studies obtain values of $r$ larger than one.
            One is the exploratory lattice calculation. This method could in principle give a reliable 
            value, if an up-to-date study would be made. The second one is the QCD sum rule estimate
            from Huang, Liu and Zhu in 1999 \cite{Huang:1999xj}. In principle this is a reliable 
            method, if it is applied properly. The calculation in \cite{Huang:1999xj}
            seems to be, however, in contradiction with the one from Colangelo and de Fazio 
            \cite{Colangelo:1996ta}. In 1996 also the method from Rosner \cite{Rosner:1996fy} 
            gave values for $r$ larger than one. This changed with new precise measurements of the
            $\Sigma_b^{(*)}$-masses. Now Rosner's methods gives a small value in accordance with the
            QCD sum rule estimate from Colangelo and de Fazio \cite{Colangelo:1996ta} and with the 
            early estimates from \cite{Guberina:1979xw}. 
            We will vary $r$ between $0.2$ (Colangelo and de Fazio) and $1.32$ ( Di Pierro, 
            Sachrajda and Michael) with a central value of $0.62$ (Rosner).
            The unclear situation with the matrix elements resulted in a broad range of  
            different theory predictions and as long as the experimental values for the
            $\Lambda_b$-lifetime were low - the HFAG average from 2003 was
            \begin{equation}
            \frac{\tau(\Lambda_b)}{\tau(B_d)}^{\rm HFAG \, 2003} =  0.798 \pm 0.034
            \end{equation}
            - there was a tendency to use preferably the larger values for $r$ in order to see,
            how far one can ``stretch'' the HQE. Estimates from that time
            \cite{Tarantino:2007nf,Gabbiani:2004tp,Franco:2002fc,Korner:2000zn,Guberina:1999bw,
            DiPierro:1999tb,Huang:1999xj,Colangelo:1996ta} read
\begin{equation}
\begin{array}{|l|l|l|}
\hline
\mbox{Year} & \mbox{Author} & \tau (\Lambda_b)/\tau (B_d)
\\
\hline
\hline
2007        & \mbox{Tarantino} &     0.88 \pm 0.05
\\
\hline
2004        & \mbox{Petrov et al.} &  0.86 \pm 0.05
\\
\hline
2002        & \mbox{Rome}          &  0.90 \pm 0.05
\\
\hline
2000        & \mbox{K{\"o}rner,Melic}   &  0.81 ... 0.92
\\
\hline 
1999        & \mbox{Guberina,Melic,Stefanic} &  0.90
\\
\hline
1999        & \mbox{diPierro, Sachrajda, Michael}   &  0.92 \pm 0.02
\\
\hline
1999        & \mbox{Huang, Liu, Zhu}  &  0.83 \pm 0.04
\\
\hline
1996        & \mbox{Colangelo, deFazio} &   >0.94
\\
\hline
\end{array}
\end{equation}
Nowadays it is clear that the low $\Lambda_b$-lifetime was a purely experimental issue. 
On the other hand the precise HQE prediction is still unknown, because we have no reliable 
calculation of the hadronic matrix elements at hand.
\\
Finally we need matrix elements of dimension six and dimension seven operators that 
are arising in mixing quantities. The status of the dimension six operators for mixing is
considerably more advanced than for the lifetime case; it is discussed in detail 
in the FLAG review \cite{Colangelo:2010et}. For the 
numerically important dimension seven contributions vacuum insertion approximation is used and
first studies with QCD sum rules have been
performed by Mannel, Pecjak and Pivovarov \cite{Mannel:2007am,Mannel:2011zza}.

\section{Status Quo of lifetimes and the HQE} 
\label{lifetime_sec4}
In this final section we update several of the lifetime predictions and compare them with the
most recent data, obtained many times at the LHC experiments.

\subsection{$B$-meson lifetimes}
The most recent theory expressions for $\tau (B^+ )/ \tau (B_s)$  and 
$\tau (B_s) / \tau (B_d)$ are given in \cite{Lenz:2011ti} (based on the calculations 
in \cite{Beneke:2002rj,Franco:2002fc,Gabbiani:2004tp,Becirevic:2001fy}).
For the charged $B$-meson we get the updated relation (including 
$\alpha_s$-corrections and $1/m_b$-corrections)
\begin{eqnarray}
\frac{\tau(B^+)}{\tau(B_d)}^{\rm HQE \, 2014} 
%\hspace{-1.4cm}
%-1
& =  & 
1 + 0.03 \left( \frac{f_{B_d}}{190.5 \, \rm MeV} \right)^2
          \left[ ( 1.0 \pm 0.2 ) B_1  
                +( 0.1 \pm 0.1 ) B_2  
          \right.
\nonumber
\\
&& \hspace{2.8cm}
\left.
               -(17.8 \pm 0.9 ) \epsilon_1 
               +( 3.9 \pm 0.2 ) \epsilon_2
               -0.26
          \right] 
\nonumber
\\
& = & 1.04^{+0.05}_{-0.01} \pm 0.02 \pm 0.01 \, .
\label{tauBp2014}
\end{eqnarray}
Here we have used the lattice values for the bag parameters from
\cite{Becirevic:2001fy}. Using all the values for the bag parameters quoted in 
Eq.(\ref{BagBmeson}), the central
value of our prediction for $\tau(B^+) / \tau (B_d)$ varies between 1.03 and 1.09.
This is indicated by the first asymmetric error  and clearly shows the urgent need for more 
profound calculations of these non-perturbative parameters.
The second error in Eq.(\ref{tauBp2014}) stems from varying the matrix elements 
of \cite{Becirevic:2001fy} in their allowed range and the third error comes 
from the renormalisation scale dependence as well as the dependence on $m_b$.
\\
Next we update also the prediction for the  $B_s$-lifetime given in \cite{Lenz:2011ti},
by including also $1/m_b^2$-corrections discussed in Eq.(\ref{delta1mb2}).
\begin{eqnarray}
\frac{\tau(B_s)}{\tau(B_d)}^{\rm HQE \, 2014} 
%\hspace{-1.4cm}
%! \! \! \! \! \! \! 
%-1
& =  & 1.003 + 
0.001 \left( \frac{f_{B_s}}{231 \, \rm MeV} \right)^2
          \left[ ( 0.77 \pm 0.10 ) B_1  
               -( 1.0 \pm 0.13 ) B_2  
          \right.
\nonumber
\\
&& \hspace{2.8cm}
\left.
               +(36 \pm 5 ) \epsilon_1 
               -(51 \pm 7) \epsilon_2
          \right] 
\nonumber
\\
& = & 1.001 \pm 0.002 \, .
\label{tauBs2014}
\end{eqnarray}
The values in Eq.(\ref{tauBp2014}) and Eq.(\ref{tauBs2014}) differ slightly
from the ones in \cite{Lenz:2011ti}, because we have used updated lattice values for the 
decay constants\footnote{We have used $f_{B_s} = 227.7$ MeV \cite{Colangelo:2010et}.} 
and we included the SU(3)-breaking of the $1/m_b^2$-correction - 
see Eq.(\ref{delta1mb2}) - for the $B_s$-lifetime, which was previously neglected. 
\
Comparing these predictions with the measurements given in Eq.(\ref{tauBexp}),
we find a perfect agreement for the $B_s$-lifetime, leaving thus only a little
space for, e.g., hidden new $B_s$-decay channels, following, e.g.,
\cite{Bobeth:2014rda,Bobeth:2011st}.
There is a slight tension in $\tau (B^+ )/ \tau (B_d)$, which, however, could 
solely be due to the unknown values of the hadronic matrix elements.
A value of, e.g., $\epsilon_1 = -0.092$ - and leaving everything else at the 
values given in  Eq.(\ref{latticebag}) - would perfectly match the current 
experimental average from Eq.(\ref{tauBexp}). Such a value of $\epsilon_2$ is within the range
of the QCD sum rule predictions \cite{Baek:1998vk,Cheng:1998ia}
shown in Eq.(\ref{BagBmeson}).
Thus, for further investigations updated lattice values for the bag parameters 
$B_1, B_2, \epsilon_1$ and 
$\epsilon_2$ are indispensable.
\\
The most recent experimental numbers for these lifetime ratios have been updated by the LHCb
Collaboration in 2014 \cite{Aaij:2014owa}.

\subsection{$b$-baryon lifetimes}
We discussed already the early stages of  the long standing puzzle related to the lifetime 
of $\Lambda_b$-baryon. After 2003 one started to find contradicting experimental values 
\cite{Aaltonen:2010pj,Aaltonen:2009zn,Abazov:2007al,Abazov:2007sf,Abulencia:2006dr,Abazov:2004bn}
- some of them 
still similarly low as the previous ones and others pointed more to a lifetime comparable to the 
one of the $B_d$-meson.

            \begin{equation}
\begin{array}{|l|l|l|c|l|}
\hline
\mbox{Year} & \mbox{Exp} & \mbox{Decay} & \tau (\Lambda_b) \left[ \mbox{ps}\right]& \tau (\Lambda_b)/\tau (B_d)
\\
\hline
\hline
2010        & \mbox{CDF} & J / \psi \Lambda   & 1.537 \pm 0.047  & 1.020 \pm 0.031
\\
\hline
2009        & \mbox{CDF} & \Lambda_c + \pi^-  & 1.401 \pm 0.058  & 0.922 \pm 0.038
\\
\hline
2007        & \mbox{D0}  & \Lambda_c \mu \nu X& 1.290 \pm 0.150  & 0.849 \pm 0.099
\\
\hline
2007        & \mbox{D0}  & J / \psi \Lambda   & 1.218 \pm 0.137  & 0.802 \pm 0.090
\\
\hline
2006        & \mbox{CDF} & J / \psi \Lambda   & 1.593 \pm 0.089  & 1.049 \pm 0.059
\\
\hline
2004        & \mbox{D0} & J / \psi \Lambda   & 1.22 \pm 0.22  & 0.87 \pm 0.17
\\
\hline
\end{array}
\end{equation}
The current HFAG average given in Eq.(\ref{taubbaryexp}) clearly rules out now the small
values of the $\Lambda_b$-lifetime.
Updating the NLO-calculation from the Rome group \cite{Ciuchini:2003ww}
and including $1/m_b$-corrections from  \cite{Gabbiani:2004tp} we get for the 
current HQE prediction
\begin{eqnarray}
\frac{\tau (\Lambda_b)}{\tau (B_d)}^{\rm HQE \, 2014} \hspace{-1cm} & = & 1 - (0.8 \pm 0.5)\%_{\frac{1}{m_b^2}} 
                                    - (4.2 \pm 3.3)\%_{\frac{1}{m_b^3}}^{\Lambda_b}
                                    - (0.0 \pm 0.5)\%_{\frac{1}{m_b^3}}^{B_d} 
                                    - (1.6 \pm 1.2)\%_{\frac{1}{m_b^4}}
\nonumber
 \\
 & = & 0.935 \pm  0.054\, ,
\label{tauLb2014}
\end{eqnarray}
where we have split up the corrections coming from the $1/m_b^2$-corrections discussed in
Eq.(\ref{delta1mb2}), the $1/m_b^3$-corrections coming from the $\Lambda_b$-matrix elements,
the $1/m_b^3$-corrections coming from the $B_d$-matrix elements
and finally  $1/m_b^4$-corrections studied in \cite{Gabbiani:2004tp}. 
The number in Eq.(\ref{tauLb2014}) is smaller than some of the previous theory predictions 
because of several reasons:
we have used updated, smaller lattice values for the decay constants, 
which gives a shift of about $+0.01$ 
in the lifetime ratio. Following our discussion of the dimension six 
matrix elements, we use three 
different determinations. Instead of using only the exploratory lattice one \cite{DiPierro:1999tb},
we also take into account 
the QCD sum rule estimate of Colangelo and de Fazio \cite{Colangelo:1996ta} and the spectroscopy
result of Rosner \cite{Rosner:1996fy}. In 1996 Rosner's method gave a large value of the matrix element.
New, precise measurements of the $\Sigma_b^{(*)}$-mass show, however, that the matrix 
element is much smaller
than originally thought. This gives 
a third enhancement factor. To obtain the final number we also scaled the 
numerical value of the $1/m_b^4$-correction with the size of $r$.
The current range of the theory prediction in Eq.(\ref{tauLb2014}) goes from 0.88 to 0.99. To reduce
this large uncertainty, new lattice calculations are necessary. In these calculations also
the penguin contractions from Fig.(\ref{fig:peng}) have to be taken into account.
\\
More recent experimental studies of the $\Lambda_b$-lifetime further strengthen the case for a
value of the lifetime ratio close to one.
The most recent and most precise measurement from LHCb gives
\cite{Aaij:2014zyy}
\begin{eqnarray}
\frac{\tau (\Lambda_b)}{\tau (B_d)}^{\rm LHCb} & = & 0.974 \pm 0.006 \pm 0.004 \, .
\end{eqnarray}
This results supersedes a previous LHCb measurement  \cite{Aaij:2013oha}.
Combined with the world average for the $B_d$-lifetime one gets
\begin{eqnarray}
\tau (\Lambda_b)^{\rm LHCb} & = & 1.479 \pm 0.009 \pm 0.010 \; \mbox{ps} \; .
\label{Lambdab_mostprecise}
\end{eqnarray}
Comparing the accuracy of these new measurements with the HFAG average given in Eq.(\ref{taubbaryexp})
shows the dramatic experimental progress.
LHCb has a further recent investigation of the $\Lambda_b$-lifetime \cite{Aaij:2014owa} - based
on different experimental techniques - and there is also a very new TeVatron (CDF)
number available \cite{Aaltonen:2014wfa}
\begin{eqnarray}
\tau (\Lambda_b)^{\rm CDF} & = & 1.565 \pm 0.035 \pm 0.020 \; \mbox{ps} 
\; .
\end{eqnarray}
All in all, now the new measurements of the $\Lambda_b$-lifetime are in nice agreement with 
the HQE result. 
This is now a very strong confirmation of the validity of the HQE and this makes also the motivation
of many of the studies
trying to explain the $\Lambda_b$-lifetime puzzle, e.g.,
 \cite{Altarelli:1996gt,Cheng:1997xba,Ito:1997qq},
invalid.
\\
In \cite{Beneke:2002rj} it was shown that the lifetime ratio of the $\Xi_b$-baryons can
be in principle be determined quite precisely, because here the above mentioned 
problems with penguin contractions
do not arise, the diagrams from Fig. \ref{fig:peng} cancel. 
Unfortunately there exists no non-perturbative determination of the 
matrix elements for $\Xi_b$-baryons. Cheng \cite{Cheng:1997xba} suggested 
to use the relation
\begin{equation}
r_{\Xi_b} = \frac{4}{3} \frac{M_{\Xi_b^*} - M_{\Xi_b}}{M_{B^*} - M_B}\; ,
\end{equation}
but there are no data available yet for the $\Xi_b^*$-mass. So, we are left with the 
possibility of
assuming that the matrix elements for $\Xi_b$ are equal to the ones of
$\Lambda_b$. In that case we can give a rough estimate for the 
expected lifetime ratio - we
update here a numerical estimate from 2008 \cite{Lenz:2008xt}.
In order to get rid of unwanted $s \to u$-transitions we define (following
\cite{Beneke:2002rj})
\begin{equation}
\frac{1}{\bar{\tau} (\Xi_b)} = \bar{\Gamma}  (\Xi_b) 
= \Gamma  (\Xi_b) -  \Gamma  (\Xi_b \to \Lambda_b + X) \, .
\end{equation}
For a numerical estimate we scan over the the results for the $\Lambda_b$-matrix 
elements obtained on the lattice by the 
study of Di Pierro, Michael and Sachrajda \cite{DiPierro:1999tb},
the QCD sum rule estimate of Colangelo and de Fazio \cite{Colangelo:1996ta} 
and the update of the spectroscopy
method of Rosner \cite{Rosner:1996fy}. 
Using also recent values for the remaining input parameters we obtain
\begin{equation}
\frac{\bar{\tau} (\Xi_b^0)}{\bar{\tau} (\Xi_b^+)}^{\rm HQE \, 2014} = 0.95 \pm 0.04 \pm 0.01 \pm ???
\, ,
\label{Xib}
\end{equation}
where the first error comes from the range of the values used for $r$, the second
denotes the remaining parametric uncertainty and $???$ stands for some unknown systematic errors, 
which comes from the approximation of the $\Xi_b$-matrix elements by the $\Lambda_b$-matrix elements.
We expect the size of these unknown systematic uncertainties not to exceed the error stemming 
from $r$, thus leading to an estimated overall error of about $\pm 0.06$. 
As soon as $\Xi_b$-matrix elements are available the ratio in Eq.(\ref{Xib}) can be 
determine more precisely than $\tau (\Lambda_b) / \tau (B_d)$.
\\
If we further approximate $\bar{\tau} (\Xi_b^0) = \tau (\Lambda_b)$ - here similar 
cancellations are expected to arise as in 
 $\tau_{B_s}/\tau_{B_d}$ - , then we arrive at the following prediction
\begin{equation}
\frac{\tau (\Lambda_b)}{\bar{\tau} (\Xi_b^+)}^{\rm HQE \, 2014}  = 0.95 \pm 0.06 \, .
\label{Xib2}
\end{equation}
From the new measurements of the LHCb Collaboration \cite{Aaij:2014esa,Aaij:2014lxa}
(see also the CDF update \cite{Aaltonen:2014wfa}), we deduce
\begin{eqnarray} 
\frac{\tau (\Xi_b^0)}{\tau (\Xi_b^+)}^{\rm LHCb\, 2014}   & = & 0.92 \pm 0.03 \; ,
\\
\frac{\tau (\Xi_b^0)}{\tau (\Lambda_b)}^{\rm LHCb\, 2014} & = & 1.006 \pm 0.021 \; ,
\\
\frac{\tau (\Lambda_b)}{\tau (\Xi_b^+)}^{\rm LHCb\, 2014} & = & 0.918 \pm 0.028 \; ,
\end{eqnarray}
which is in perfect agreement with the predictions above in Eq.(\ref{Xib})
and  Eq.(\ref{Xib2}), within the current uncertainties.

\subsection{$D$-meson lifetimes}
In \cite{Lenz:2013aua} the NLO-QCD corrections for the $D$-meson lifetimes were completed.
Including $1/m_c$-corrections as well as some assumptions about the hadronic matrix elements
one obtains
\begin{eqnarray}
\frac{\tau (D^+)}{\tau (D^0)}^{\rm HQE \, 2013} & = & 
{2.2 \pm 0.4^{(\rm hadronic)}}^{+0.03(\rm scale)}_{-0.07} \; ,
\\ 
\frac{\tau (D^+_s)}{\tau (D^0)}^{\rm HQE \, 2013} & = & 
{1.19 \pm 0.12^{(\rm hadronic)}}^{+0.04(\rm scale)}_{-0.04} \; ,
\end{eqnarray}
being very close to the experimental values shown in Eq.(\ref{tauDexp}). Therefore this
result seems to indicate that one might apply the HQE also to lifetimes of $D$-mesons,
but definite conclusions cannot not be drawn without a reliable non-perturbative determination
of the hadronic matrix elements, which is currently missing.

\subsection{Mixing quantities}
The current status of mixing quantities, both in the $B$- and the $D$-system, 
was very recently reviewed in \cite{Lenz:2014nka}. The arising set of observables
allows for model-independent searches for new physics effects in mixing, 
see e.g. \cite{Lenz:2010gu,Lenz:2012az}.
We discuss here only the decay rate differences $\Delta \Gamma_s$, because this provided
one of the strongest proofs of the HQE.
The HQE prediction - based on the 
NLO-QCD corrections \cite{Beneke:1998sy,Beneke:2003az,Ciuchini:2003ww,Lenz:2006hd}
and
sub-leading HQE corrections \cite{Beneke:1996gn,Dighe:2001gc}
gave in 2011 \cite{Lenz:2011ti}
\begin{equation}
\Delta \Gamma_s^{\rm HQE \, 2011} = \left(0.087 \pm 0.021 \right)\, {\rm ps}^{-1} \, .
\label{DGsHQE}
\end{equation}
$\Delta \Gamma_s$ was measured for the first time in 2012 by the LHCb
Collaboration \cite{LHCb:2012}. The current average from HFAG \cite{HFAG}
reads
\begin{equation}
\Delta \Gamma_s^{\rm Exp.} = \left( 0.091 \pm 0.09 \right)\, {\rm ps}^{-1} \, ,
\label{DGsExp}
\end{equation}
it includes the measurements from LHCb \cite{Aaij:2013oba,Aaij:2014zsa}, 
ATLAS \cite{Aad:2012kba,Aad:2014cqa},
CMS \cite{CMS:2014jxa}, CDF \cite{Aaltonen:2012ie} and D0 \cite{Abazov:2011ry}.
Experiment and theory agree perfectly for $\Delta \Gamma_s$, excluding thus huge 
violations of quark hadron duality. The new question is now: how precisely does the HQE work?
The experimental uncertainty will be reduced in future,
while the larger theory uncertainty is dominated from unknown matrix elements of 
dimension seven operators, see \cite{Lenz:2006hd,Lenz:2011ti}. Here a first lattice 
investigation or a continuation of the QCD sum rule study in \cite{Mannel:2007am,Mannel:2011zza}
would be very welcome.

\section{Conclusion}
\label{lifetime_sec5}
We have started this review by  giving a very basic introduction into lifetimes 
of weakly decaying particles, followed by a detailed discussion of the individual 
terms appearing in the HQE. Next we focused on the historical development of the 
theory, which we summarise briefly as:
early investigations of the  HQE are based on the work by Voloshin and Shifman 
\cite{Khoze:1983yp, Shifman:1984wx} in the early 1980s. 
A real systematic expansion was only possible
after some conceptual issues have been solved in 1992 by Bigi and Uraltsev  
\cite{Bigi:1991ir}, which was proven in 1998 by Beneke et al. \cite{Beneke:1998sy} 
in an explicit calculation.   
The HQE in its present form was developed in 1992 by Bigi, Uraltsev and Vainshtein
\cite{Bigi:1992su} and about the same time by Blok and Shifman 
\cite{Blok:1992hw,Blok:1992he}.
For semi-leptonic decays the absence of $1/m_q$-corrections was already shown in 1990
by  Chay, Georgi and Grinstein \cite{Chay:1990da} and by Luke \cite{Luke:1990eg}.
\\
Since 1992 several discrepancies were arising, that shed some doubt on the 
validity of the HQE: inclusive non-leptonic decays (in particular predictions 
for the semi-leptonic branching ratio and the missing charm puzzle) 
and the $\Lambda_b$-lifetime were two prominent examples.
We have discussed in detail, how all these issues were resolved.
For the semi-leptonic branching ratio NLO-QCD corrections including finite charm-quark mass 
effect were crucial. The remaining small difference, see Eq.(\ref{Bsltheo}) 
vs. Eq.(\ref{Bslexp}) is probably due to unknown NNLO-QCD effects. The problem
of the $\Lambda_b$-lifetime was experimentally solved in the last months.
One of the most convincing tests of the HQE was, however, the measurement of $\Delta \Gamma_s$
from 2012 onwards - see Eq.(\ref{DGsExp}) - in perfect agreement with the prediction stemming from early
2011 - see Eq.(\ref{DGsHQE}).
\\
Thus, the theory in whose development Kolya played such a crucial role, has just now
passed numerous non-trivial tests and its validity holds beyond any doubt. This makes also
the motivation for looking for some modification  of the HQE, see e.g. 
\cite{Falk:1994hm,Altarelli:1996gt,Cheng:1997xba,Ito:1997qq} 
invalid.
The new question is now: how precise is the HQE?
This question is not only of academic interest, but it has practical consequences in 
searches for new physics. The quantification of a statistical significance of a possible
discrepancy depends strongly on the intrinsic uncertainty of the HQE. Hence further 
studies in that direction are crucial.
As a starting point for such an endeavour we have updated several theory predictions
for lifetime ratios
\begin{eqnarray}
\frac{\tau (B^+)}{\tau (B_d)}^{\rm HQE \, 2014} 
& = & 1.04^{+0.07}_{-0.03} \; ,
\\
\frac{\tau (B_s)}{\tau (B_d)}^{\rm HQE \, 2014} 
& = & 1.001 \pm 0.002 \; ,
\\
\frac{\tau (\Lambda_b)}{\tau (B_d)}^{\rm HQE \, 2014} 
& = & 0.935 \pm 0.054 \; ,
\\
\frac{\bar{\tau} (\Xi_b^0)}{\bar{\tau} (\Xi_b^+)}^{\rm HQE \, 2014} 
& = & 0.95 \pm 0.06 \; .
\end{eqnarray}
In order to see, how these predictions could be further improved, we compare for 
different observables
what components of the theory prediction are currently known. 
\begin{displaymath}
\begin{array}{|c||c|c|c|c|c|}
\hline
& \frac{\tau(B+)}{\tau (B_d)} & \Gamma_{12} & \frac{\tau(\Lambda_b)}{\tau (B_d)} 
& \frac{\tau(D^+)}{\tau (D_0)}& \frac{\bar{\tau} (\Xi_b^0)}{\bar{\tau} (\Xi_b^+)}
\\
\hline
\hline
\Gamma_3^{(0)}           & +        & + & + & + & +
\\
\Gamma_3^{(1)}           & +        & + & 0 & + & +
\\
\Gamma_3^{(2)}           & -        & - & - & - & -
\\
\langle \Gamma_3 \rangle & 0        & + & 0 & - & -
\\
\hline
\Gamma_4^{(0)}           & +        & + & + & + & +
\\
\Gamma_4^{(1)}           & -        & - & - & - & -
\\
\langle \Gamma_4 \rangle & -        & 0 & - & - & -
\\
\hline
\Gamma_5^{(0)}           & -        & + & + & -& +
\\
\langle \Gamma_5 \rangle & -        & - & - & - & -
\\
\hline
\end{array}
\end{displaymath}
For all these observables the LO-QCD term $\Gamma_3^{(0)}$ and also
the NLO-QCD corrections $\Gamma_3^{(1)}$ are known. For the $\Lambda_b$-baryon, however, 
a part of the NLO-QCD calculation is still missing. NNLO-QCD corrections - denoted
by $\Gamma_3^{(2)}$ - have not been 
calculated for any of these observables; a first step for $\Gamma_{12}$ has been done in
\cite{Asatrian:2012tp}.
The biggest problem are currently the non-perturbative matrix elements. Concerning 
the dimension 6 term $\langle \Gamma_3 \rangle$ we have only 
for $\Gamma_{12}$ several independent lattice calculations.
For $\tau (B^+) / \tau (B_d)$ the latest lattice number stems from 2001 \cite{Becirevic:2001fy},
for $\tau (\Lambda_b) / \tau (B_d)$ we have only an exploratory lattice study from 
1999 \cite{DiPierro:1999tb}
and for the $D$-meson lifetimes we have no lattice investigations at all. 
For the $b$-hadrons also several QCD sum rule determinations of these matrix elements
are available \cite{Baek:1998vk,Cheng:1998ia,Colangelo:1996ta,Huang:1999xj}.
\\
Concerning the power suppressed $1/m_b$ corrections, we see that the LO-QCD term $\Gamma_4^{(0)}$
is known for all observables and $\Gamma_5^{(0)}$ is also known for some of the observables. 
The matrix elements of the dimension seven operators, $\langle \Gamma_4 \rangle$, have been determined 
by vacuum insertion approximation - a first step of a QCD sum rule calculation for $\Delta \Gamma_s$ has
be done in \cite{Mannel:2011zza,Mannel:2007am}. 
\\
For all lifetime ratios the uncertainty due to the unknown matrix elements of the
dimension six operators
is dominant. For $\Delta \Gamma_s$ these operators have already been determined by 
several groups and thus the dominant uncertainty stems now from $\Gamma_4$. Here 
a full non-perturbative determination of the matrix elements of the dimension seven operators
would be very desirable, as well as calculation of the corresponding NLO-QCD corrections, denoted by 
$\Gamma_4^{(1)}$.
Increasing the precision of the HQE will also help in shrinking the allowed space for new 
physics effects in
tree-level decays \cite{Brod:2014bfa}, a topic that has also profound implications for other
branches of flavour physics. 
\\
Kolya left us a very promising but also challenging legacy, which might in the end 
provide the way to identify new physics in the flavour sector.

\section*{Acknowledgements}
I would like to thank 
Martin Beneke, 
Christoph Bobeth, 
Markus Bobrowski, 
Gerhard Buchalla, 
Christoph Greub,
Uli Haisch, 
Fabian Krinner, 
Uli Nierste, 
Ben Pecjak, 
Thomas Rauh, 
Johann Riedl, 
J{\"u}rgen Rohrwild 
and  
Gilberto Tetlalmatzi-Xolocotzi
for collaborating on topics related to the determination of lifetime predictions.
Valerie Khoze, Ikaros Bigi, Jonathan Rosner and Mikhail Voloshin for helpful 
comments on the history of lifetimes.
Many thanks to Gilberto Tetlalmatzi-Xolocotzi for producing most of the Feynman diagrams,
to Fabian Krinner and Thomas Rauh for numerical updates of the coefficient $c_3$ and to 
Paolo Gambino,
Valery Khoze, Uli Nierste, Thomas Rauh, Jonathan Rosner and Gilberto Tetlalmatzi-Xolocotzi 
for proof-reading.

\bibliographystyle{ws-rv-van}
\bibliography{ws-rv-sample}

%\printindex[aindx]                 % to print author index
%\printindex                         % to print subject index
\end{document}